\pdfoutput=1

\documentclass[11pt]{article}

\usepackage[final]{acl}

\usepackage{times}
\usepackage{amsthm}
\usepackage{xcolor}
\usepackage{latexsym}
\usepackage{subfig}
\usepackage{multirow}
\usepackage{amssymb}
\usepackage{array} 
\usepackage{booktabs}
\usepackage{algorithm}
\usepackage{mdframed}
\usepackage{algorithmic}
\usepackage[T1]{fontenc}

\usepackage[utf8]{inputenc}

\usepackage{microtype}
\usepackage{amsmath}
\theoremstyle{plain} %
\newtheorem{theorem}{Theorem} %

\usepackage{inconsolata}

\usepackage{graphicx}

\newenvironment{ack}{\section*{Acknowledgements}}{}

\title{Enhancing LLM-Based Social Bot via an Adversarial Learning Framework}

\author{%
  Fanqi Kong$^{2,1}$, Xiaoyuan Zhang$^{2,1}$, Xinyu Chen$^{2}$, Yaodong Yang$^{2}$, \\ \textbf{Song-Chun Zhu$^{1,2,3,4}$, Xue Feng$^{1}$\thanks{Corresponding Author.}}\\ \\ 
  $^1$State Key Laboratory of General Artificial Intelligence, BIGAI \\ 
  $^2$Peking University \quad \quad $^3$Tsinghua University \\
  $^4$PKU-WUHAN Institute for Artificial Intelligence \\
    \texttt{kfq20@stu.pku.edu.cn, fengxue@bigai.ai}
}

\begin{document}
\maketitle
\begin{abstract}

Developing Large Language Model (LLM) agents that exhibit human-like behavior, encompassing not only individual heterogeneity rooted in unique user profiles but also adaptive response to socially connected neighbors, is a significant research challenge. Social media platforms, with their diverse user data and explicit social structures, provide an ideal testbed for such investigations. This paper introduces EvoBot, an \textbf{Evo}lving LLM-based social \textbf{Bot} that significantly enhances human-like generative capabilities through a novel adversarial learning framework. EvoBot is initialized by Supervised Fine-Tuning (SFT) on representative data from social media and then iteratively refines its generation of sophisticated, human-like content via Direct Preference Optimization (DPO). This refinement is guided by feedback from a co-adapting \textbf{Detector} which concurrently improves its ability to distinguish EvoBot from humans, thereby creating an increasingly challenging learning environment for EvoBot. Experiments demonstrate that EvoBot generates content aligned with diverse user profiles, increasingly bypassing the co-adapting Detector through human-like expression. Moreover, it exhibits strong social responsiveness, more accurately modeling real-world opinion dynamics and information spread in multi-agent simulations. The framework also yields a more robust Detector, underscoring its broader utility for both advanced agent development and related detection tasks. The code is available at \url{https://github.com/kfq20/EvoBot}.
\end{abstract}

\section{Introduction}

A key aspiration in Large Language Models (LLMs) is to create autonomous agents that exhibit human-like behavior, moving beyond mere textual fluency to embody the richness of human interaction.  Human-likeness in this context is profoundly multifaceted. It centrally involves individual heterogeneity, where an agent's expressions and actions are authentically rooted in unique personal characteristics, historical context, and specific profile attributes \cite{putnam2000bowling, tajfel1979integrative}. Simultaneously, it requires sophisticated social intelligence, reflecting how individuals dynamically perceive and respond to their socially connected neighbors and the broader social structure, leading to complex emergent phenomena such as opinion dynamics \cite{chuang2023simulating, ma2024potential}, social influence \cite{abbas2020effect, peng2016social} and information spread (like rumors, social-disease contagion) \cite{chopra2024limits, bauch2013social, feng2018voluntary, feng2019dynamic}. Endowing LLM-based agents with both this deep individual distinctiveness and adaptive social responsiveness remains a significant research challenge.

To instill multifaceted human-likeness in agents, learning environments must reflect real-world complexity. Social media platforms offer a uniquely rich setting for this: they combine detailed individual data (e.g., user profiles, posts, interaction histories) with complex social graphs (e.g., follower networks, communities). This confluence enables LLM agents to model both personalized expression and socially adaptive behavior \cite{gao2023s, gao2024large, yang2024oasis, kong2025siv}. However, pre-trained LLMs often fail to capture the full range of human preferences, particularly those of marginalized communities, resulting in generic or biased outputs \cite{cheng2023compost, chakrabortymaxmin, he2024community}. Prompt engineering offers a lightweight method to guide model behavior without retraining, but its effectiveness comes with trade-offs: detailed prompts can improve authenticity, yet often reduce scalability and efficiency, echoing challenges seen in large-scale simulations. Fine-tuning provides stronger alignment with specific behaviors and personalities \cite{shao2023character, ge2024scaling}, but it typically relies on human-labeled or high-quality synthetic data, which can introduce bias and limit generalizability \cite{chopra2024limits, williams2023epidemic, rafailov2024direct, ouyang2022training, josifoski2023exploiting, deng2023rephrase}. This gap highlights the need for more effective learning paradigms, leading to our core research question:

\begin{center}
    \textit{How can LLMs learn from social media data to generate more human-like content?}
\end{center}

To address this question, we propose a novel EvoBot (\textbf{Evo}lving LLM-based social \textbf{Bot}) through an adversarial learning framework, which is realized by leveraging the task of social bot detection, a well-established research area on social media \cite{feng2022twibot, feng2021twibot, cresci2015fame, cresci2019better}. Specifically, a co-adapting \textbf{Detector} serves as an increasingly discerning `adversary' providing dynamic feedback throughout the training of EvoBot, in contrast to traditional detectors trained on static datasets or not designed to address LLM-based bots \cite{yang2020scalable, feng2021botrgcn, dialektakis2022caleb, wu2019detecting}.

EvoBot’s learning begins with Supervised Fine-Tuning (SFT), where it is prompted with summarized user account information and details of their network neighbors to learn to generate tweet outputs that mimic genuine human communication patterns. Subsequently, EvoBot enters an iterative adversarial loop to refine its tweet generation for improved sophistication and authenticity via Direct Preference Optimization (DPO) \cite{rafailov2024direct}. The crucial preference signals for DPO are derived from the Detector's classifications of EvoBot's generated tweets, guiding it towards outputs the Detector finds more human-like. The Detector, in turn, is regularly updated with EvoBot's evolving tweet outputs, ensuring a continually escalating challenge that propels EvoBot's learning.

The efficacy of our adversarial framework is substantiated by theoretical analysis and extensive empirical results, demonstrating EvoBot's advancements in achieving multifaceted human-likeness. At the individual level, EvoBot learns to generate content that reflects diverse user profiles and achieves a high degree of human-like expression, evidenced by its increasing success in bypassing the Detector. 
Furthermore, at the group level, EvoBot demonstrates its capacity for human-like social responsiveness, outperforming baseline models by more accurately replicating real-world group opinion and information spread. A key driver of EvoBot’s improvements is the dynamic adversarial learning environment shaped by the co-adapting Detector, which continually retrains on EvoBot’s evolving outputs. 
This process also improves the Detector’s classification performance and generalization, making it a progressively stringent and adaptive benchmark for EvoBot's continued learning and development.

\begin{figure*}
    \centering
    \includegraphics[width=0.99\linewidth]{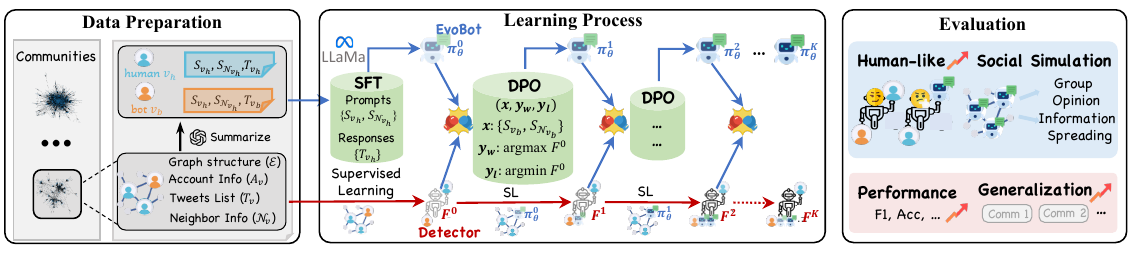}
    \caption{Overview of the EvoBot Framework, detailing its three core stages:
\textbf{(1) Data Preparation:} Extraction of user and interaction data from social media communities. \textbf{(2) Learning Process:} EvoBot is initialized through SFT on human data. It then iteratively refines its human-like content generation via DPO, using feedback from a co-adapting Detector that serves as an evolving evaluative benchmark. \textbf{(3) Evaluation:} Assessment of EvoBot’s enhanced multifaceted human-likeness, both at an individual level (e.g., through interaction with the Detector) and at a group level (via its ability to drive realistic group opinion and information spread simulations), alongside the Detector’s resulting classification performance and generalization.}
    \label{fig:main}
\end{figure*}

\section{Methodology}
The learning process of EvoBot includes two phases. \textbf{1)} SFT is conducted on real human data to pre-train EvoBot on the expressive habits, linguistic styles, and contextual preferences of community members. \textbf{2)} Adversarial learning is used, with both the EvoBot and the Detector iteratively trained. EvoBot's objective is to generate tweets that are most likely to be classified as human by the Detector, while the Detector aims to improve its accuracy in distinguishing between bots and humans. The following parts provide a problem formulation and a detailed description of all modules. Figure \ref{fig:main} provides an overview of this framework. And the learning process is detailed in Algorithm \ref{alg:evobot}.

\subsection{Problem Formulation} \label{notation}
The social media dataset is modeled as a tuple \((\mathcal{V}, \{A_v\}, \{T_v\}, \mathcal{E}) \), where \( \mathcal{V} \) is the set of nodes, representing users, divided into two classes \( \mathcal{H} \) (humans) and \( \mathcal{B} \) (bots), i.e., \( \mathcal{V} = \mathcal{H} \cup \mathcal{B} \).  
\( \mathcal{E} \) is the set of directed edges, where \( (u, v) \in \mathcal{E} \) indicates that user \( u \) follows user \( v \).
Each user \( v \in \mathcal{V} \) is associated with two types of attributes:  
\textbf{Account Information} \( A_v = \{a_1, a_2, \dots, a_m\} \), which includes account features on Twitter, such as account creation time, number of followers, user description, and so on.  
\textbf{Tweets} \( T_v = \{t_1, t_2, \dots, t_n\} \), which represents a set of tweets posted by \( v \).  

The adversarial learning proceeds for \(K\) rounds. In the \(k\)-th round, \textbf{EvoBot}, represented by \(\pi^k_{\theta}\), generates tweets for a target user by integrating both the user's and their neighbors' information. Specifically, for user \(v\), we use GPT-4o-mini \cite{hurst2024gpt} (\(M_{\text{sum}}\)) to condense their account information \(A_v\) and historical tweets \(T_v\) into a concise summary \(S_v = M_{\text{sum}}(A_v, T_v)\), which forms the first input. Similarly, the neighbor information is summarized as \(S_{\mathcal{N}_v} = M_{\text{sum}}(A_{\mathcal{N}_v}, T_{\mathcal{N}_v})\). These, along with a task instruction \(\mathcal{I}\), guide EvoBot to generate tweets that align with the target user's profile and fit naturally into the community. The tweets are then generated as \(T_v \propto \pi_{\theta}^k(T_v | \mathcal{I}, S_v, S_{\mathcal{N}_v})\).

\textbf{Detector} in the \(k\)-th round is defined as \(F^{k}=\sum_{j=0}^{k}w^jf^j\), where \(f:(\mathbf{A}, \mathbf{T}, \mathcal{E}) \rightarrow \mathbf{p}\) is the classifier trained in each round. Note that \(F^0=f^0\) is the base detector trained on the original dataset. \(\mathbf{A}=\{A_1, A_2, \dots, A_N\}\),\(\mathbf{T}=\{T_1, T_2, \dots, T_N\}\) represent account information and tweets for all \(N\) users, respectively. \(\mathbf{p} = [p_1, p_2, \dots, p_N]\) is the vector of probabilities, where \(p_v \in [0, 1]\) is the probability that user \(v\) is classified as a human.

\subsection{EvoBot} \label{Sec: EvoBot}

\textbf{Supervised Fine-Tuning} 
The SFT dataset is constructed by randomly selecting a subset of human users \( \mathcal{H}_{\text{SFT}} \subseteq  \mathcal{H} \). For each \( v_h \in \mathcal{H}_{\text{SFT}} \), the prompt is \((\mathcal{I}, S_{v_h}, S_{\mathcal{N}_{v_h}})\). The reference
 response is the \( l \) tweets \( T_{v_h}=\{t_{v_h,1}, t_{v_h,2}, \dots, t_{v_h,l}\} \) sampled from the user's historical tweets \( T_{v_h} \).

The objective of SFT is to minimize the discrepancy between the tweets generated by the base model of EvoBot \(\pi_{\theta}^0\) and the reference responses. This is achieved by optimizing the negative log-likelihood loss: \(\mathcal{L}_{\text{SFT}} = -\frac{1}{|\mathcal{H}_{\text{SFT}}|} \sum\limits_{{v_h} \in \mathcal{H}_{\text{SFT}}} \log \pi_{\theta}^0(T_{v_h} | \mathcal{I}, S_{v_h}, S_{\mathcal{N}_{v_h}}).\)

\textbf{Adversarial Learning with Detector} 
EvoBot is trained to generate tweets that evade detection as bot-generated. A naive approach would be to use the Detector’s output—the probability of being classified as human—as a scalar reward in reinforcement learning \cite{chen2024off, zhang2025differentiable, kong2024learning}. However, this faces issues like reward sparsity and unstable gradient estimation, leading to inefficient and suboptimal training \cite{cao2023reinforcement, zhang2024overcoming}. So we use DPO—a fine-tuning approach that directly uses the preference ordering in the data rather than training an additional reward model \cite{rafailov2024direct}. 

Specifically, in the \(k\)-th round of adversarial learning, \(N\) bot users \(\{v_{b_i} \in \mathcal{B}, i=1, \dots, N\}\) are randomly sampled with replacement, and EvoBot generates \(C\) candidate responses \(\{T_{v_{b_i}, c}\}_{c=1, \dots, C}\) for each \(v_{b_i} \). The Detector \(F^k\) then evaluates each candidate while keeping all other users' information fixed, calculating the probability that \(v_{b_i}\) is human for each response \(T_{v_{b_i}, c}\), denoted as \(F^k_{v_{b_i}, c}\).

To construct the DPO dataset \(\mathcal{D}_{\text{DPO}}=\{x^i, y_w^i, y_l^i\}_{i=1}^N\), we let \(x^i=(\mathcal{I}, S_{v_{b_i}}, S_{\mathcal{N}_{v_{b_i}}})\), \(y_w^i = \arg\max_c F^{k}_{v_{b_i},c}\), \(y_l^i = \arg\min_c F^{k}_{v_{b_i},c}\), where \(x^i\) is the input context, and \(y_w^i\) and \(y_l^i\) are the tweets with the highest and lowest probabilities of being classified as human, respectively.
The loss function is \resizebox{\columnwidth}{!}{\(\mathcal{L}_{\text{DPO}}=-\mathbb{E}_{x^i, y_w^i, y_{l}^i}\left[\log \sigma\left(\beta \log \frac{\pi^k_{\theta}\left(y_w^i \mid x^i\right)}{\pi^{k-1}_{\theta}\left(y_w^i \mid x^i\right)} -\beta \log \frac{\pi^k_{\theta}\left(y_l^i \mid x^i\right)}{\pi^{k-1}_{\theta}\left(y_l^i \mid x^i\right)}\right)\right],\)} where \(\sigma\) is the sigmoid function, and \(\beta\) is a hyperparameter controlling the deviation from the \(k-1\)'s version of EvoBot.

\subsection{Detector}
\textbf{Features extraction} 
Our detector employs a feature extraction approach inspired by \citep{feng2021botrgcn}. More specifically, the classifier \( f: (\mathbf{A}, \mathbf{T}, \mathcal{E}) \rightarrow \mathbf{p} \) takes as input account information \( \mathbf{A} \), tweets \( \mathbf{T} \), and the graph structure \( \mathcal{E} \). The account information \( A_v \) includes numerical properties such as account creation time and number of followers, which are normalized for balanced scaling, as well as categorical properties like user description and verified status, represented using one-hot encoding for binary interpretability. The textual data in \( T_v \) is embedded by RoBERTa \cite{liu2019roberta} to capture semantic content. These features are processed through separate linear layers with LeakyReLU activations and then combined into a unified embedding. To incorporate the graph structure, we use an RGCN layer that aggregates relational information from the graph \( \mathcal{E} \) based on the relation types. The resulting embeddings pass through fully connected layers with dropout regularization, producing a binary classification output that predicts whether a user is a bot or a human.

\textbf{Supervised Learning} In the \(k\)-th round of adversarial training, to obtain Detector \(F^k\), all bot tweets in the dataset are replaced with outputs generated by EvoBot \(\pi^{k-1}_{\theta}\) from the previous round. This modified dataset is then used to train the classifier \(f^k\) via supervised learning, using a cross-entropy loss to maximize classification accuracy.

\subsection{Theoretical Analysis} \label{sec: theoretical analysis}
In this subsection, we provide a theoretical analysis for our method from a more general view. We assume that data on social platforms can be represented as \({(\mathbf{x}, \mathbf{y})}\), where \(\mathbf{x}\) denotes various user attributes, such as age, gender, occupation, interests, etc., sampled from the marginal distribution \(q(\cdot)\) of the entire community. Meanwhile, \(\mathbf{y} \sim \pi_{\mathcal{H}}(\cdot | \mathbf{x})\) represents the user's activities on the platform, such as posting tweets, retweeting, and liking, where \(\pi_{\mathcal{H}}\) is the decision model of real humans in the community. Similarly, \(\pi_\theta\) denotes EvoBot's model.  

Detector is trained to maximize the probability of correctly classifying real and fake samples:  
\begin{equation}
\begin{split}
    F = & \arg \max \mathbb{E}_{\mathbf{x} \sim q(\cdot), \mathbf{y} \sim \pi_{\mathcal{H}}(\cdot | \mathbf{x})}[\log F(\mathbf{x}, \mathbf{y})] \\
    + & \mathbb{E}_{\mathbf{x}' \sim q'(\cdot), \mathbf{y}' \sim \pi_{\theta}(\cdot |\mathbf{x}')}[\log(1 - F(\mathbf{x}', \mathbf{y}'))]
\end{split}
\label{eq:detector}
\end{equation}
Here, the inputs \(\mathbf{x}' \sim q'(\cdot)\) for EvoBot are distinguished from \(q(\cdot)\), indicating that the input information received by EvoBot may come from a different distribution than the input received by real humans.

Considering the construction method of our DPO dataset in Section \ref{Sec: EvoBot} and referring to \citet{rafailov2024direct}, the optimization objective of EvoBot is:
\begin{equation}
\small
    \begin{split}
        \pi_\theta & =  \arg\min  \mathbb{E}_{\mathbf{x}' \sim q'(\cdot), \mathbf{y}' \sim \pi_\theta(\cdot |\mathbf{x}')}[1-\log F(\mathbf{x}', \mathbf{y}')] \\
        + & \beta \mathbb{E}_{\mathbf{x} \sim q(\cdot), \mathbf{x}' \sim q'(\cdot)}[KL(\pi_{\mathcal{H}}(\cdot|\mathbf{x}) \| \pi_\theta(\cdot|\mathbf{x}'))].
    \end{split}
    \label{eq: generator}
\end{equation}

\begin{theorem}
    If \(q'(\mathbf{x}) = q(\mathbf{x})\), then under the iterative training process of the detector and generator with the optimization objective (\ref{eq:detector}) and (\ref{eq: generator}), the global optimum is achieved when \(\pi_\theta = \pi_{\mathcal{H}}\).
    \label{thm: 1}
\end{theorem}

The proof is provided in Appendix \ref{appendix: theorems}.
\section{Experiment Setup}
\subsection{Dataset}
We use TwiBot-22 \cite{feng2022twibot}, a graph-based Twitter dataset that includes one million users, nearly one hundred million tweets, and various relational data. In this dataset, we represent users as nodes and model follower-followee relationships as directed edges in a graph. Given the impracticality of training EvoBot on the entire dataset due to its size and complexity, we divide the network into smaller, more manageable communities using the \textit{Louvain} community detection method \cite{blondel2008fast}, identifying 12 highly connected and representative communities. These communities exhibit diverse network topologies (e.g., star-shaped, mesh-like structures), support multiple languages, and focus on a variety of topics. See Appendix \ref{appendix: dataset} for data details and the preprocessing.

\subsection{Models and Trainings}
EvoBot is based on Llama-2-7b-chat \cite{touvron2023llama}. For fine-tuning, we use the \texttt{transformers} and \texttt{trl} libraries to implement SFT and DPO. And we apply low-rank adaptation (LoRA) \cite{hu2021lora} using the \texttt{peft} library.

Training and inference are performed on 8 NVIDIA RTX 3090 GPUs, with each community requiring approximately 10 hours. EvoBot runs \(K=4\) iterations for adversarial learning, using 1024-sample datasets for both SFT and DPO. Detector is trained with an 8:1:1 split for training, validation, and test sets, with performance evaluated on the test set.
Model architectures and training hyperparameters are detailed in the Appendix \ref{appdix: training}.

\subsection{Simulation Framework}\label{sec: sim}
We use the open-source social media simulation framework HiSim \cite{mou2024unveiling} to analyze the response dynamics of EvoBot and baseline models as users react to trigger events, focusing on group opinion and information spread. Since EvoBot is designed for tweet generation, we simplify the simulation by excluding user actions like likes and retweets. EvoBot directly controls human accounts from the dataset to simulate user behavior. At each step, the input includes the prompt from Section \ref{Sec: EvoBot}, the current trigger event, recent past events, and the latest tweets from followed users.

\section{Results}
This section presents two sets of experiments. First, we assess EvoBot's enhanced generation of individually authentic content, alongside the co-adapting Detector that serves as an increasingly stringent benchmark. Second, we evaluate EvoBot's capacity for human-like social responsiveness at a group level through simulations of real-world group opinion and information spread.
\subsection{Individual-Level Human-Likeness}
\begin{figure}[ht]
  \includegraphics[width=\columnwidth]{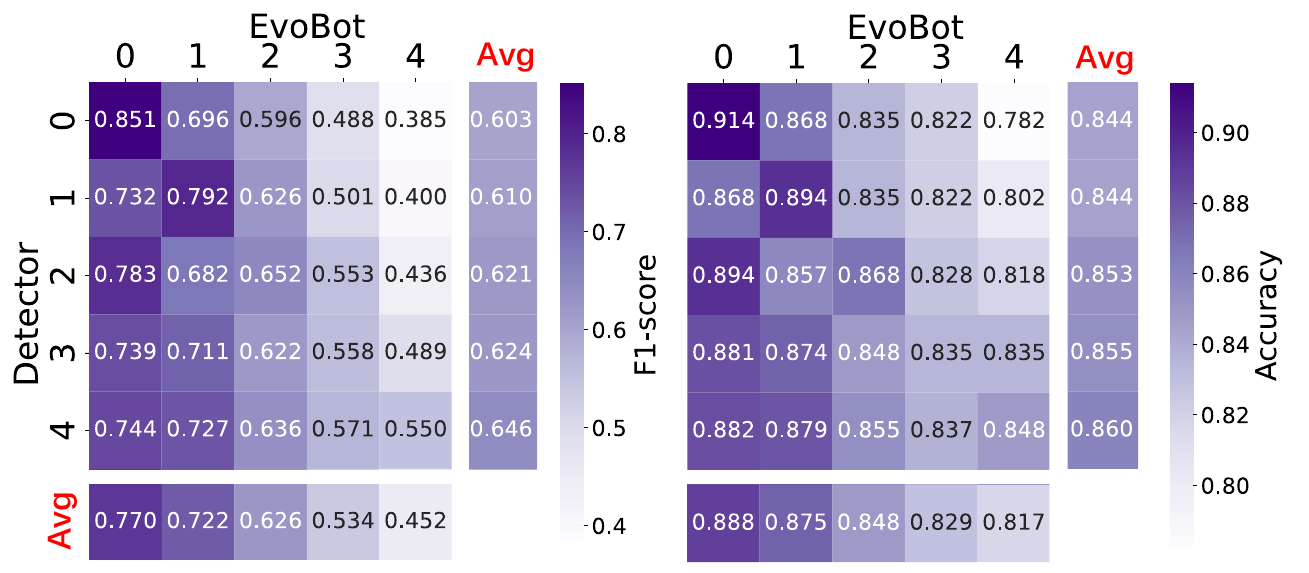}
  \caption{Classification performance across training iterations. Left: F1-score; right: accuracy. Rows indicate Detector versions; columns indicate EvoBot versions.}
  \label{fig:adv results}
  \vspace{-0.3cm}
\end{figure}

To evaluate the impact of adversarial training on both EvoBot and the Detector, Figure \ref{fig:adv results} presents the classification performance in terms of F1-score and Accuracy (averaged across 12 different communities) across various training iterations. The use of both metrics is crucial due to the class imbalance between positive (human) and negative (bot) samples in the dataset. In the heatmaps, the cell at Detector \(F^i\) (row \(i\)) and EvoBot \(\pi^j_{\theta}\) (column \(j\)) quantifies the result of \(F^i\) classifying bots that have been replaced with outputs from \(\pi^j_{\theta}\).

These average metrics highlight the co-adaptive learning dynamics: the `Avg' raw for EvoBot shows a steady decline in average F1-score (from 0.770 for $\pi_{\theta}^{0}$ to 0.452 for $\pi_{\theta}^{4}$), indicating that later versions become increasingly evasive; meanwhile, the `Avg' column for the Detector reveals progressively stronger performance across all EvoBot versions as training advances. However, performance is not strictly monotonic across all pairings—for instance, Detector $F^2$ may underperform $F^1$ against an earlier EvoBot such as $\pi_{\theta}^1$, possibly due to overfitting on weaker or less diverse early outputs. Nevertheless, the overall trend reflects co-adaptive progress: EvoBot becomes more evasive and natural, while the Detector grows more robust—a pattern that also holds when using Llama-3-8B-Instruct \cite{llama3modelcard} as EvoBot’s base, as shown in Appendix \ref{appdx:generator}, Figure \ref{fig:llama3 results}.
Furthermore, the adversarial training exhibits signs of convergence, as indicated by the diagonal F1-scores tracking Detector $F^k$ against its counterpart EvoBot $\pi_{\theta}^k$ (0.851 → 0.792 → 0.652 → 0.558 → 0.550), which stabilize from $k=3$ onward.

\subsubsection{EvoBot's Performance} 

\begin{table*}[t]
    \centering
    \resizebox{\textwidth}{!}{
    \begin{tabular}{l|l|ccccccc}
    \toprule
         \textbf{Detector} & \textbf{Metric} $\downarrow$ & \textbf{Origin}  & \textbf{GAN} & \textbf{Llama2-7b} & \textbf{GPT-4o-mini} & \textbf{w/o Adv} & \textbf{w/o SFT} & \textbf{Ours} \\ 
         \midrule
         \multirow{2}{*}{\textbf{RGCN}} & Accuracy & $\underline{0.827\pm 0.067}$& $0.853 \pm 0.088$& $0.849\pm 0.050$& $0.851\pm 0.044$& $0.833\pm 0.070$& $0.834\pm 0.049$ & $\mathbf{0.805\pm 0.084}$\\
          & F1-score & $0.455 \pm 0.045$& $0.584 \pm 0.164$& $0.497 \pm 0.051$& $0.458 \pm 0.041$& $0.454 \pm 0.038$& $\underline{0.449 \pm 0.052}$ & $\mathbf{0.393 \pm 0.036}$\\
          \midrule
          \multirow{2}{*}{\textbf{GAT}} & Accuracy & $0.836 \pm 0.040$  & $0.865 \pm 0.046$ & $0.847 \pm 0.037$ & $0.834 \pm 0.050$ & $\underline{0.818 \pm 0.063}$ & $0.844 \pm 0.045$ & $\mathbf{0.788 \pm 0.092}$ \\
           & F1-score & $0.424 \pm 0.046$  & $0.515 \pm 0.089$ & $0.474 \pm 0.041$ & $\underline{0.407 \pm 0.032}$ & $0.440 \pm 0.008$ & $0.440 \pm 0.051$ & $\mathbf{0.355 \pm 0.031}$ \\
         \bottomrule
    \end{tabular}}
    \caption{Accuracy and F1-score of different generators using bot RGCN and GAT detectors. The detectors are trained on the original dataset. A smaller value indicates stronger ability of the generator to evade detection.}
    \label{tab:generator results}
\end{table*}

We compare the generative capabilities of the final version of EvoBot \(\pi^4_{\theta}\) (\textbf{Ours}) and other models in generating human-like tweets under the Detector \(F^0\) trained on the original dataset. We include six baselines: \textbf{(1) Origin}: Bots from the original dataset; \textbf{(2) GAN}: Implemented using the PyTorch-GAN open-source repository \cite{LinderNoren_PyTorchGAN_2017}. Due to the non-differentiable nature of discrete text, the generator is trained to produce vectors matching the dimensionality of tweet embeddings from RoBERTa; \textbf{(3) Llama2-7b; (4) GPT-4o-mini}: The two pre-trained LLMs use the same prompts and generation parameters as EvoBot; \textbf{(5) w/o ADV}: This ablation removes the adversarial learning process by training for only one iteration. To maintain a constant total amount of training data, the DPO dataset is scaled to \(N=KN\). \textbf{(6) w/o SFT}: This ablation removes the SFT phase. 
Additionally, to further assess the generative capacity of different models, we replace the original RGCN-based Detector with a \textbf{GAT} model \cite{velivckovic2017graph}.

Table \ref{tab:generator results} presents the Detector's classification performance for different generators. Smaller values indicate stronger generator performance, as the generated content becomes more difficult to distinguish from tweets by human users. EvoBot consistently outperforms other models, effectively evading the Detectors and achieving the lowest classification accuracy and F1-score. In contrast, GAN struggles to capture meaningful language features in the embedding space, making it highly detectable by the Detector and resulting in the poorest performance. Moreover, the generated vectors fail to decode into coherent, natural language. Both Llama and GPT perform worse than EvoBot. The two ablation studies highlight the importance of both SFT and adversarial training. The above findings hold across both detector architectures, demonstrating consistent results across different setups.

We assess EvoBot's output diversity using n-gram metrics (Dist-1, -2, -3) and Shannon Entropy (Table \ref{tab:evobot_diversity}), where higher scores generally indicate richer, more varied language. We use the Wilcoxon test to assess EvoBot’s diversity gains, with the largest and most significant improvement from v0 to v1 and more gradual refinements in later versions (Table \ref{tab:diversity_significance}). While diversity tends to increase as EvoBot evolves, several metrics (Dist-1, Dist-3, Shannon Entropy) peak at EvoBot version 3, with version 4 values remaining comparably high. This suggests that after several adversarial iterations, EvoBot may approach a point of stabilization or near-optimal performance for these specific diversity aspects, with later refinements potentially yielding marginal changes or focusing on other unmeasured qualities of human-like generation. Importantly, the continued improvement in Dist-2 through version 4 indicates ongoing enhancement in other facets of textual variety. Illustrative examples of EvoBot's generated content are provided in Appendix Figures \ref{fig:generator_example}, \ref{fig:eg covid}, and \ref{fig:eg info spread}.

\begin{table}[h!]
\centering
\resizebox{0.5\textwidth}{!}{
\begin{tabular}{lccccc}
\toprule
\textbf{EvoBot Version} & \textbf{0} & \textbf{1} & \textbf{2} & \textbf{3} & \textbf{4} \\ \midrule
Dist-1 & 0.1835 & 0.1871 & 0.1939 & \textbf{0.2129} & \(\underline{0.2100}\) \\
Dist-2 & 0.7440 & 0.7511 & 0.7545 & \(\underline{0.7574}\) & \textbf{0.7627} \\
Dist-3 & 0.9294 & 0.9342 & 0.9331 & \textbf{0.9479} & \(\underline{0.9451}\) \\
Shannon Entropy & 11.5703 & 11.6591 & 11.6310 & \textbf{11.6753} & \(\underline{11.6662}\) \\ \bottomrule
\end{tabular}
}
\caption{Output diversity for different versions of EvoBot. Higher values indicate greater diversity.}
\label{tab:evobot_diversity}
\vspace{-0.3cm}
\end{table}

We also analyze the output length and stylistic markers to understand how EvoBot's outputs change qualitatively. As Table \ref{tab:length_stylistic} shows, adversarial training guides EvoBot towards a more concise and human-like length. While the vanilla LLM (`Llama`) produces lengthy tweets, the SFT phase (v0) dramatically reduces tweet length, and later adversarial phases (v3, v4) make outputs even shorter, aligning with real human tweets. This trend is statistically significant (Mann-Whitney U test, p $<$ 0.001), indicating the framework successfully optimizes for conciseness.

Our analysis of stylistic markers confirms a two-phase learning dynamic. The initial SFT phase corrects the overly frequent, bot-like usage of emojis and hashtags from the base LLM. The subsequent adversarial phases then compel EvoBot to use these features within a more natural, fluctuating range, demonstrating a more nuanced, context-dependent understanding of human-like expression.

\begin{table*}[t]
\centering
\resizebox{\textwidth}{!}{
\begin{tabular}{llcccccccc}
\toprule
\textbf{Metric} & \textbf{Type} & \textbf{Llama} & \textbf{v0 (SFT)} & \textbf{v1} & \textbf{v2} & \textbf{v3} & \textbf{v4} & \textbf{Human} \\
\midrule
\multirow{2}{*}{\textbf{Output Length}} & Mean Char Length & 237.97 $\pm$ 83.60 & 252.04 $\pm$ 147.93 & 243.75 $\pm$ 124.52 & 240.34 $\pm$ 132.17 & 160.22 $\pm$ 74.54 & 175.99 $\pm$ 88.41 & 94.24 $\pm$ 63.23 \\
& Mean Word Count & 36.04 $\pm$ 12.95 & 36.98 $\pm$ 8.97 & 38.14 $\pm$ 6.96 & 37.65 $\pm$ 7.14 & 24.99 $\pm$ 6.70 & 25.99 $\pm$ 7.04 & 15.86 $\pm$ 10.69 \\
\midrule
\multirow{3}{*}{\textbf{Stylistic Usage}} & Emoji Usage & 52.2\% & 14.1\% & 18.4\% & 16.8\% & 12.6\% & 14.4\% & 17.1\% \\
& Hashtag Usage & 69.5\% & 30.3\% & 12.4\% & 13.9\% & 4.9\% & 14.9\% & 18.6\% \\
& Mention Usage & 15.0\% & 6.6\% & 4.8\% & 6.3\% & 5.4\% & 7.0\% & 10.8\% \\
\bottomrule
\end{tabular}
}
\caption{Analysis of EvoBot's output length and stylistic features compared to the vanilla LLM and real humans. Note that all length shifts from one version to the next are statistically significant (p $<$ 0.001) unless marked as non-significant (ns) in the rebuttal, and the effect size of these shifts are negligible where noted.}
\label{tab:length_stylistic}
\end{table*}

\subsubsection{Detector's Performance}
\begin{table*}[t]
    \centering
    \resizebox{\textwidth}{!}{
    \begin{tabular}{l|ccccccccc}
    \toprule
         \textbf{Metric} $\uparrow$ & \textbf{Origin}  & \textbf{Random} & \textbf{Exp.1} & \textbf{Exp.5} & \textbf{Greedy} & \textbf{GAT} & \textbf{w/o RGCN} & \textbf{w/o $T$} & \textbf{Ours} \\ 
         \midrule
          Accuracy & $0.827\pm 0.067$& $0.224 \pm 0.031$ &$\underline{0.882\pm 0.067}$ & $0.880\pm 0.025$ & $0.875\pm 0.033$ & $0.868\pm 0.042$ &$0.849\pm 0.075$ &$0.829\pm 0.065$&$\mathbf{0.892\pm 0.053}$\\
          F1-score & $0.424 \pm 0.046$ & $0.169\pm 0.031$ & $\underline{0.550\pm 0.040}$ & $0.526\pm 0.039$ &$0.457\pm 0.014$ &$0.500\pm 0.075$ &$0.432\pm 0.060$ & $0.350\pm 0.071$ &$\mathbf{0.561\pm 0.042}$\\
         \bottomrule
    \end{tabular}}
    \caption{Accuracy and F1-score of different detectors evaluated on the original dataset. A larger value indicates a stronger ability of the detector to distinguish between human and bot.}
    \label{tab:detector results}
    \vspace{-0.3cm}
\end{table*}
In this section, we evaluate the performance of the final version of Detector \(F^4\) through comparisons and ablation studies, using classification accuracy and F1-score on the original dataset. We include 7 baselines. First, we examine different strategies for selecting the classifier weights \( w^j \) in the Detector. Our method, \textbf{Ours}, uses a uniform weighting strategy where \( w^j = \frac{1}{k} \), assigning equal weight to each classifier. We compare this with the \textbf{(1) Greedy} approach, which assigns \( w^j = 1 \) to the most recent classifier and \(w^j=0\) otherwise, and the \textbf{(2) Exp.1} and \textbf{(3) Exp.5} strategies, where \( w^j = e^{-\alpha (k-j)} \), with \( \alpha =0.1\) and \(\alpha=0.5\), respectively. Additionally, we evaluate \textbf{(4) GAT}, the previously discussed GAT-based model, and perform two ablation studies: \textbf{(5) w/o RGCN}, where the RGCN structure is removed, and \textbf{(6) w/o \(\mathbf{T}\)}, where tweet features are excluded from the input. Finally, we include the \textbf{(7) Random} baseline, where labels are assigned randomly, as a lower bound for performance.

The classification performance, shown in Table \ref{tab:detector results}, leads to several key conclusions. \textbf{First}, both \textit{Ours} and \textit{Exp} outperform \textit{Greedy}, highlighting the crucial role of EvoBot in the iterative training process. This suggests that earlier versions of EvoBot still benefit the Detector’s learning. The performance improvement is primarily driven by data augmentation— as more EvoBot versions are added, the diversity and quantity of bot-generated tweets increase, enhancing supervised learning. \textbf{Second}, the results of \textit{GAT} and \textit{w/o RGCN} emphasize the importance of the RGCN structure, which plays a vital role in capturing relational data and structural information within the graph. \textbf{Third}, the \textit{w/o \(T\)} and \textit{Random} results demonstrate that tweet content is essential for effective classification.

\begin{figure}[ht]
    \centering
    \includegraphics[width=0.49\textwidth]{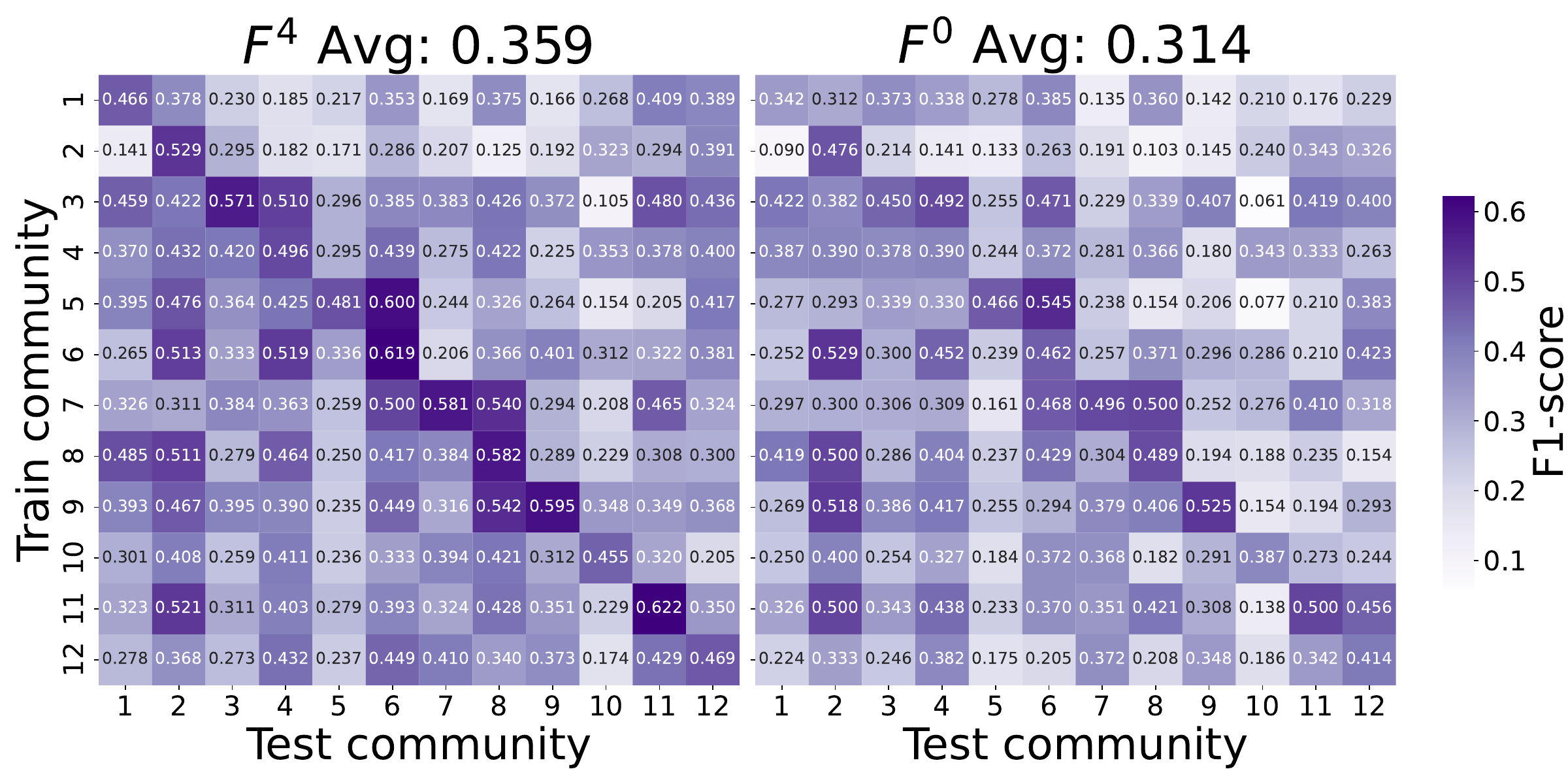}
    \caption{The generalization ability of detectors trained on one community and tested on others. Left shows results for the final version of Detector $F^4$, while right is for the Detector $F^0$ trained on the original dataset.}
    \label{fig:generalization}
\end{figure}

Next, we evaluate the Detector's generalization ability by training it on data from a single community and testing it on the test sets of all communities. Figure \ref{fig:generalization} shows that training the Detector with EvoBot-generated data improves its cross-community generalization compared to a Detector trained solely on the original dataset. Using F1-score as the evaluation metric, the results indicate that the adversarially trained Detector outperforms the one trained on the original data, demonstrating better generalization across communities.

To further evaluate the generalization capabilities of our evolved detector ($F^4$), we compare its performance against the original detector ($F^0$) on two external datasets: Cresci-15 \cite{cresci2015fame} and TwiBot-20 \cite{feng2021twibot}. The results are presented in Table \ref{tab:external_detector_performance}, suggesting the adversarial training process improves the detector's generalization to unseen datasets.

\begin{table}[h!]
\centering
\resizebox{0.48\textwidth}{!}{
\begin{tabular}{l|cc|cc}
\toprule
\multirow{2}{*}{\textbf{Detector}} & \multicolumn{2}{c|}{\textbf{Cresci-15}} & \multicolumn{2}{c}{\textbf{TwiBot-20}} \\
 & \textbf{Acc} & \textbf{F1} & \textbf{Acc} & \textbf{F1} \\ \midrule
Evolved detector ($F^4$) & 0.4367 & 0.2850 & 0.5254 & 0.3425 \\
Original detector ($F^0$) & 0.3705 & 0.1476 & 0.4976 & 0.2369 \\ \bottomrule
\end{tabular}
}
\caption{Performance Comparison on External Datasets}
\label{tab:external_detector_performance}
\vspace{-0.3cm}
\end{table}

\subsection{Group-Level Human-Likeness}
\subsubsection{Group Opinion}

We simulate group opinion during two major events: the \textbf{COVID-19 pandemic} and the \textbf{Russia-Ukraine Conflict}. For COVID-19, we select one key event per month from January 2020 to March 2022, simulating over \(T = 27\) steps. For the Russia-Ukraine Conflict, we choose one significant event per day from February 24 to March 13, 2022, simulating over \(T = 18\) steps.

We use the sentiment analysis model of \citet{barbieri2020tweeteval} to score each user's post on a scale from -1 (negative) to +1 (positive) at each time step. The opinion of user \(i\) at time \(t\) is denoted as \(O_{i,t}\). For each time step, we compute the mean \(\bar{O_t}\) and standard deviation \(\sigma_t\) of the opinions across all users:
\(
\bar{O_t} = \frac{1}{N} \sum_{i=1}^{N} O_{i,t}\), \(\sigma_t = \sqrt{\frac{1}{N} \sum_{i=1}^{N} (O_{i,t} - \bar{O_t})^2}.
\)

We report four metrics of the results: the average group opinion across all time steps, \(Mean = \frac{1}{T}\sum_{t=1}^T\bar{O_t}\), which reflects the overall opinion trend of the group; the average standard deviation of group opinion, \(Std = \frac{1}{T}\sum_{t=1}^T\sigma_t\), capturing the diversity of opinions within the group; the average bias, \(\Delta_{Bias} = \frac{1}{T} \sum_{t=1}^{T} \left| \bar{O_t} - \bar{O}_{real,t} \right|\), between the simulated and real group opinions; and the average difference in opinion diversity, \(\Delta_{Div} = \frac{1}{T} \sum_{t=1}^{T} \left| \sigma_t - \sigma_{real,t} \right|\), assessing how well the simulation replicates the variance in group opinions. Here, \(\bar{O}_{real,t}\) and \(\sigma_{real,t}\) are derived from the real data during the corresponding real-time period.

We use four baselines: \textbf{Llama2-7b}, \textbf{GPT-4o-mini}, and two well-known Agent-Based Models (ABMs): \textbf{Bounded Confidence (BC)} model \cite{deffuant2000mixing} and \textbf{Lorenz} model \cite{lorenz2021individual}. The BC model updates agents' opinions only when a received message meets a predefined confidence threshold. The Lorenz model accounts for mechanisms like contagion, assimilation, motivated cognition, attitude formation, polarity, and source credibility to simulate the evolution of individual opinions. Both models are initialized with the real community network structure and user opinions, then iterated until convergence.

Table \ref{tab:group attitude} presents the results. For the real-world dataset, group opinion on COVID-19 tends to be relatively neutral, while the Russia-Ukraine Conflict elicits more negative sentiment, including expressions of fear and condemnation. In both cases, individual opinions vary widely, as reflected in the high standard deviation of sentiment scores. BC and Lorenz models, which often result in opinion convergence or polarization, are limited by their fixed, rule-based interactions that oversimplify the dynamics of opinion formation and cannot fully represent complex, changing real-world events. In contrast, LLMs like GPT and Llama generate more diverse content but tend to produce overly generic responses. When discussing complex topics, they often resort to simplified, advocacy-oriented content, missing the range of real-world sentiments. Among all models, EvoBot exhibits the smallest \(\Delta_{Bias}\) and \(\Delta_{Div}\), indicating that its generated opinions align most closely with real-world data in terms of group bias and individual diversity. The statistical significance of these improvements is further confirmed by a Mann-Whitney U test, as detailed in Table \ref{tab:simulation_significance}. We also conduct robustness checks to confirm the stability of our simulation results across different random seeds and communities, with the full analysis presented in Table \ref{tab:simulation_stability}. See Appendix \ref{appendix: group opinion} for more details.

\begin{table}[ht]
    \centering
    \small
    \resizebox{0.48\textwidth}{!}{
    \setlength{\tabcolsep}{2pt} %
    \newcommand{\textBF}[1]{\textbf{\text{#1}}}
    
    \begin{tabular}{l|*{4}{w{r}{0.9cm}}|*{4}{w{r}{0.9cm}}}
    \toprule
    \multirow{2}{*}{\textbf{Method}} & \multicolumn{4}{c|}{\textbf{COVID-19}} & \multicolumn{4}{c}{\textbf{Russian-Ukrainian Conflict}} \\
    \cmidrule(lr){2-5} \cmidrule(lr){6-9}
    & \textit{Mean} & \textit{Std} & \(\Delta_{Bias}\) & \(\Delta_{Div}\) & \textit{Mean} & \textit{Std} & \(\Delta_{Bias}\) & \(\Delta_{Div}\) \\
    \midrule
    Real & -0.017 & 0.472 & \multicolumn{1}{c}{/} & \multicolumn{1}{c|}{/} & -0.239 & 0.670 & \multicolumn{1}{c}{/} & \multicolumn{1}{c}{/} \\
    BC   & -0.041 & 0.389 & 0.089 & 0.112 & -0.304 & 0.104 & 0.104 & 0.554 \\
    Lorenz & 0.084 & 0.725 & 0.107 & 0.264 & -0.811 & 0.105 & 0.572 & 0.565 \\
    Llama & -0.053 & 0.368 & 0.098 & 0.105 & -0.324 & 0.405 & 0.202 & 0.265 \\
    GPT & 0.032 & 0.342 & 0.081 & 0.083 & -0.256 & 0.435 & 0.135 & 0.238 \\
    EvoBot & 0.010 & 0.428 & \textbf{0.072} & \textbf{0.052} & -0.237 & 0.480 & \textbf{0.101} & \textbf{0.194} \\
    \bottomrule
    \end{tabular}
    }
    \caption{Simulation results for group opinion.}
    \label{tab:group attitude}
    \vspace{-0.3cm}
\end{table}

\subsubsection{Information Spread}

Information spread in social networks shapes public discourse, influences opinions, and determines how events gain attention. To study this, we focus on a baseball community discussing the Los Angeles Rams' Super Bowl LVI victory. Using keyword matching, we identify users engaging in these discussions and track participation over time. For the simulation, we select the first 30 users to post about the event as the initial sources of information, with only these users having access to the information at the start. Information then spreads through the real network structure, where users receive updates via tweets from accounts they follow.

\begin{figure}[ht]
    \centering
    \includegraphics[width=0.48\textwidth]{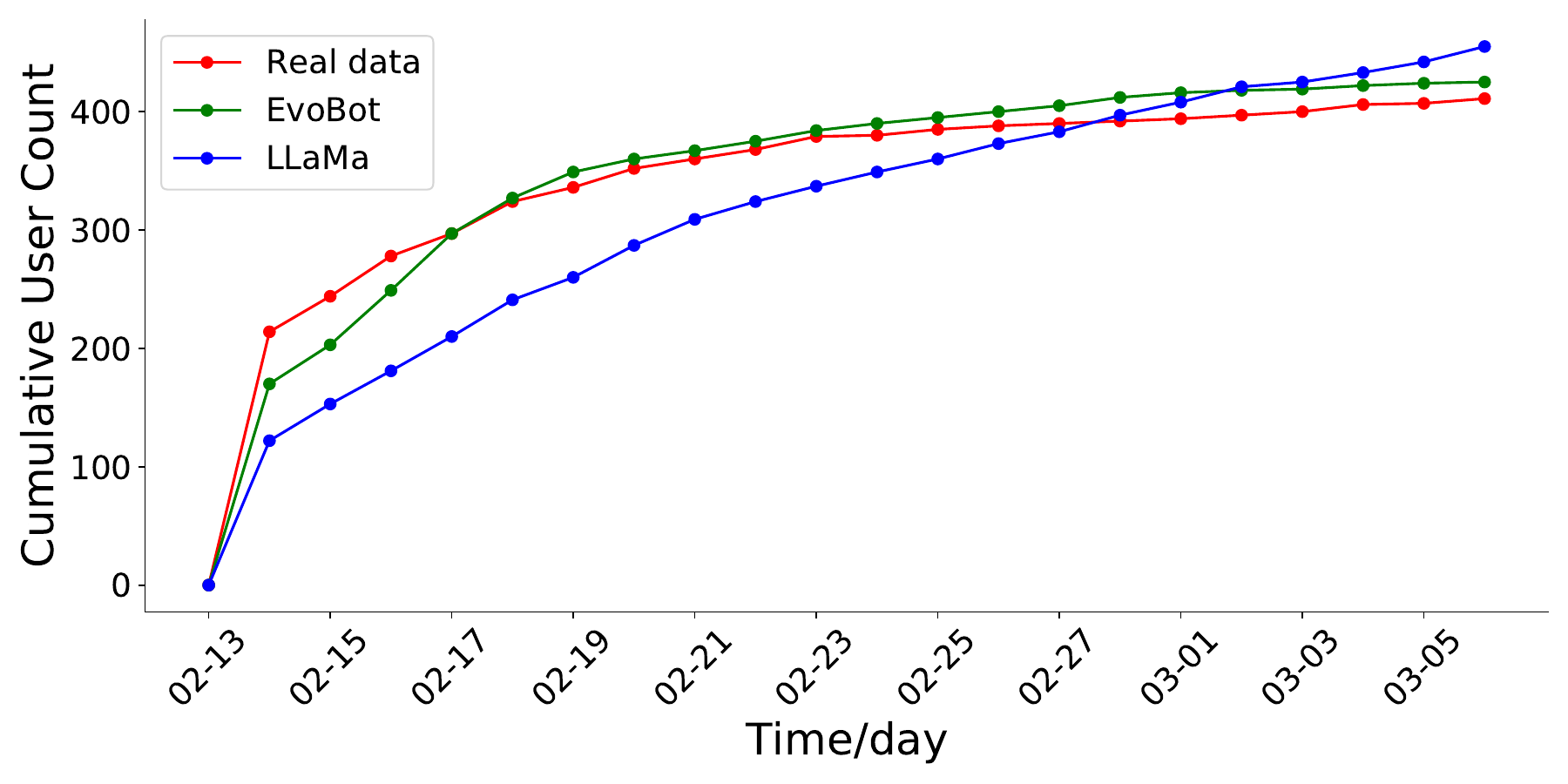}
    \caption{Cumulative author count discussing the Los Angeles Rams' Super Bowl LVI victory over time, highlighting the growth of online buzz every 24 hours.}
    \label{fig:ram}
    \vspace{-0.2cm}
\end{figure}

The results are shown in Figure \ref{fig:ram}. Compared to Llama, EvoBot’s simulation results align better with real-world information spread, successfully replicating the phenomenon of rapid initial spread followed by a gradual slowdown. EvoBot’s more direct and concise responses contribute to this effectiveness, facilitating faster and broader dissemination of information, as demonstrated in Figure \ref{fig:eg info spread}. However, since we restrict users to receiving information solely through the posts of others, while in the real world, people have many other ways of obtaining information, there is still some deviation, especially in the early stage.

\section{Related Works}
\textbf{LLM-based Agents in Social Simulation.} Recent studies have explored the use of LLMs as autonomous agents in social simulation, categorizing them into individual, scenario, and society-level simulations \cite{mou2024individual}. Individual-level studies focus on modeling specific personas or demographic groups to analyze behavioral patterns \cite{shao2023character, shen2023roleeval, frisch2024llm}. Scenario-based simulations involve structured interactions among multiple LLM-driven agents to tackle domain-specific tasks, such as software development \cite{qian2023communicative, hong2023metagpt}, question answering \cite{du2023improving}, and judicial decision-making \cite{he2024simucourt}. At the societal level, multi-agent simulations have been employed to examine emergent social behaviors \cite{park2023generative, yang2024oasis}, including opinion dynamics \cite{chuang2023simulating} and macroeconomic trends \cite{li2024econagent}.
EvoBot stands out by using a learning-based approach for agents to adapt and improve, unlike most methods that rely on prompt engineering.

\textbf{Adversarial Learning.} Adversarial learning has been successfully applied in traditional NLP tasks like text generation \cite{yu2017seqgan, li2017adversarial}, and more recently in social bot detection: GANs have been used to generate synthetic bot samples to address class imbalance \cite{wu2019detecting, wu2020using, dialektakis2022caleb}, but these methods often struggle with detecting evolved bots that adapt to bypass detection systems. \citet{cresci2020decade} introduced a proactive detection method using genetic algorithms, while \citet{jan2020throwing} proposed a GAN-based framework with two generators to detect advanced bot variants. With the rise of LLMs, AI-generated text detection has become more challenging \cite{wu2025survey}, though some adversarial methods \cite{hu2023radar, koike2024outfox} have improved detection accuracy. Unlike these methods, which focus solely on detection, EvoBot's dual focus on both generation and detection makes it a powerful tool in the ongoing arms race between AI creators and detection systems.

\section{Conclusion}

This paper introduces EvoBot, an LLM-based social bot enhanced through a novel adversarial learning framework. EvoBot learns to generate content that reflects authentic individual traits and socially adaptive behavior, refining its outputs through interaction with a co-adapting Detector that creates an increasingly challenging environment. Experiments show that EvoBot achieves  more human-like expression and better captures social dynamics in group opinion and information spread simulations. The adversarial process also produces a more capable and adaptive Detector. Our approach offers a promising path toward developing nuanced, context-aware social agents for dynamic settings like social media, emphasizing the utility of adversarial learning with a domain-grounded evaluator.

\section*{Limitations}
There are several limitations to the current approach. First, the Detector’s fixed training parameters during adversarial learning could benefit from automated tuning to balance performance and overfitting. Second, limited resources constrained training to a smaller dataset and fewer epochs, which may affect generalization. Lastly, while the framework performs well in a controlled setting, maintaining stability, adaptability, and robustness at real-world scale remains a major challenge. These limitations point to future directions for improving EvoBot’s resilience and flexibility.

\section*{Ethics Statement}
We collect and process data from the publicly available TwiBot-22 dataset in compliance with its original terms. We remove personally identifiable information (e.g., URLs, phone numbers, emails) from tweets using keyword matching, and anonymize all user names. However, like most LLMs, EvoBot may generate harmful content. Therefore, we implement strict review procedures to ensure the model is used only for research purposes. EvoBot shows promise in generating realistic content, but its ethical implications must be considered. The ability to create human-like text could be misused for disinformation or manipulation. While our current work focuses on a foundational learning framework, we acknowledge the importance of technical safeguards for responsible deployment. In future work, we will explore integrating specific defense mechanisms, such as content watermarking schemes, to embed a traceable signature into generated text. We will also investigate real-time filtering and algorithmic auditing designs to mitigate misuse and ensure transparency. These measures will be crucial for establishing ethical guidelines and regulatory frameworks to mitigate risks.

EvoBot shows promise in generating realistic content, but its ethical implications must be considered. The ability to create human-like text could be misused for disinformation or manipulation. Future work should focus on establishing safeguards and transparency measures to ensure responsible use, along with ethical guidelines and regulatory frameworks to mitigate risks.

\section*{Broader Impact}
EvoBot could drive advancements in AI-human interaction and enhance applications like personalized communication and social media management. Additionally, the development of a more generalized Detector with stronger generalization capabilities will play a crucial role in distinguishing human from machine-generated content, ensuring the responsible deployment of such technologies.

\begin{ack}
    This work is supported by the \textit{Wuhan East Lake High-Tech Development Zone (Optics Valley of China, OVC) National Comprehensive Experimental Base for Governance of Intelligent Society.}
\end{ack}

\bibliography{custom}

\appendix

\section{Proof of Theorems in Section \ref{sec: theoretical analysis}} \label{appendix: theorems}
The proof of Theorem \ref{thm: 1}:

\textit{Proof.} Considering \(q(\mathbf{x})=q'(\mathbf{x})\), the maximization objective in (\ref{eq:detector}) when generator \(\pi_{\theta}\) is fixed can be written as:
\begin{equation*}
    \begin{split}
        V(F) &=\mathbb{E}_{\mathbf{y} \sim \pi_{\mathcal{H}}(\cdot | \mathbf{x}), \mathbf{x} \sim q(\cdot)}[\log F(\mathbf{x}, \mathbf{y})] \\
        &+ \mathbb{E}_{\mathbf{y}' \sim \pi_{\theta}(\cdot |\mathbf{x}'), \mathbf{x}' \sim q'(\cdot)}[\log(1 - F(\mathbf{x}', \mathbf{y}'))] \\
        &= \int_\mathbf{x}q(\mathbf{x})\int_\mathbf{y} \pi_{\mathcal{H}}(\mathbf{y} | \mathbf{x})\log F(\mathbf{x}, \mathbf{y})d\mathbf{y}d\mathbf{x} \\
        &+ \int_\mathbf{x}q(\mathbf{x})\int_\mathbf{y} \pi_{\theta}(\mathbf{y} | \mathbf{x})\log(1 - F(\mathbf{x}, \mathbf{y}))d\mathbf{y}d\mathbf{x} \\
        &= \int_\mathbf{x}q(\mathbf{x})\int_\mathbf{y} \pi_{\mathcal{H}}(\mathbf{y} | \mathbf{x})\log F(\mathbf{x}, \mathbf{y}) \\
        &+ \pi_{\theta}(\mathbf{y} | \mathbf{x})\log(1 - F(\mathbf{x}, \mathbf{y})) d\mathbf{y}d\mathbf{x}
    \end{split}
\end{equation*}
Let \(L(F) = \pi_{\mathcal{H}}(\mathbf{y} | \mathbf{x})\log F(\mathbf{x}, \mathbf{y}) + \pi_{\theta}(\mathbf{y} | \mathbf{x})\log(1 - F(\mathbf{x}, \mathbf{y}))\), the derivative of \(L\) with respect to \(F\) is:
\[
L'(F) = \frac{dL}{dF} = \frac{\pi_{\mathcal{H}}}{F} - \frac{\pi_{\theta}}{1-F}
\]
To find the maximum of \(L\), we set \(L'(F) = 0\) and get the optimal detector \(F^*(\mathbf{x},\mathbf{y})\):
\[
F^*(\mathbf{x},\mathbf{y})=\frac{\pi_{\mathcal{H}}(\mathbf{y}|\mathbf{x})}{\pi_{\mathcal{H}}(\mathbf{y}|\mathbf{x}) + \pi_{\theta}(\mathbf{y}|\mathbf{x})}
\]
It can be observed that for \(\pi_{\mathcal{H}} = \pi_{\theta}\), \(F^*(\mathbf{x},\mathbf{y}) = \frac{1}{2}\), meaning that the detector is unable to distinguish between samples generated by the generator and real samples, and can only classify them randomly with a probability of 0.5.

Assuming the detector has reached its optimal state \(F^*(\mathbf{x},\mathbf{y})\) , the generator's minimization objective can be written as:
\begin{equation*}
    \begin{split}
        & V(\pi_{\theta})= \mathbb{E}_{\mathbf{x} \sim q(\cdot), \mathbf{y} \sim \pi_\theta(\cdot |\mathbf{x})}[1-\log F^*(\mathbf{x}, \mathbf{y})] \\
        &+ \beta \mathbb{E}_{\mathbf{x} \sim q(\cdot)}[KL(\pi_{\mathcal{H}}(\cdot|\mathbf{x}) \| \pi_\theta(\cdot|\mathbf{x}))] \\
        &= -\log(2) \\
        &+ \mathbb{E}_{\mathbf{x} \sim q(\cdot), \mathbf{y} \sim \pi_\theta(\cdot |\mathbf{x})}[\log \frac{2\pi_{\theta}(\mathbf{y}|\mathbf{x})}{\pi_{\mathcal{H}}(\mathbf{y}|\mathbf{x}) + \pi_{\theta}(\mathbf{y}|\mathbf{x})}] \\
        &+\beta \mathbb{E}_{\mathbf{x} \sim q(\cdot)}[KL(\pi_{\mathcal{H}}(\cdot|\mathbf{x}) \| \pi_\theta(\cdot|\mathbf{x}))] \\ 
        &= -\log(2) \\
        &+ \int_\mathbf{x}q(\mathbf{x})\int_\mathbf{y} \pi_\theta(\mathbf{y}|\mathbf{x})\log \frac{\pi_{\mathcal{H}}(\mathbf{y}|\mathbf{x})}{\frac{\pi_{\theta}(\mathbf{y}|\mathbf{x}) + \pi_{\theta}(\mathbf{y}|\mathbf{x})}{2}}\\
        &+ \beta \mathbb{E}_{\mathbf{x} \sim q(\cdot)}[KL(\pi_{\mathcal{H}}(\cdot|\mathbf{x}) \| \pi_\theta(\cdot|\mathbf{x}))] \\
        &= -\log(2) + \mathbb{E}_{\mathbf{x} \sim q(\cdot)}\left[KL\left(\pi_{\theta}\| \frac{\pi_{\mathcal{H}}+\pi_{\theta}}{2}\right)\right]\\
        &+ \beta \mathbb{E}_{\mathbf{x} \sim q(\cdot)}[KL(\pi_{\mathcal{H}}(\cdot|\mathbf{x}) \| \pi_\theta(\cdot|\mathbf{x}))] 
    \end{split}
\end{equation*}

Since the KL divergence is always non-negative and achieves zero only when the distributions being compared are identical, the two KL terms in the objective function will both be minimized (i.e., equal to zero) when  \(\pi_{\theta} = \pi_{\mathcal{H}}\). Therefore, the global minimum of the objective function is achieved when \( \pi_{\theta} = \pi_{\mathcal{H}}\) , as both KL divergence terms vanish, leading to the optimal solution. \qed

\label{sec:appendix}

\section{Data Details and Preprocessing}\label{appendix: dataset}
Our study utilizes the TwiBot-22 dataset, which is publicly available under the MIT License. The dataset was originally designed for bot detection research, and we ensure that our use aligns with this intended purpose. We do not repurpose or distribute the dataset beyond research contexts. Additionally, any derivative data created in this study is used solely for academic research and follows the original access conditions.

\begin{figure}[ht]
    \centering
    \includegraphics[width=0.98\linewidth]{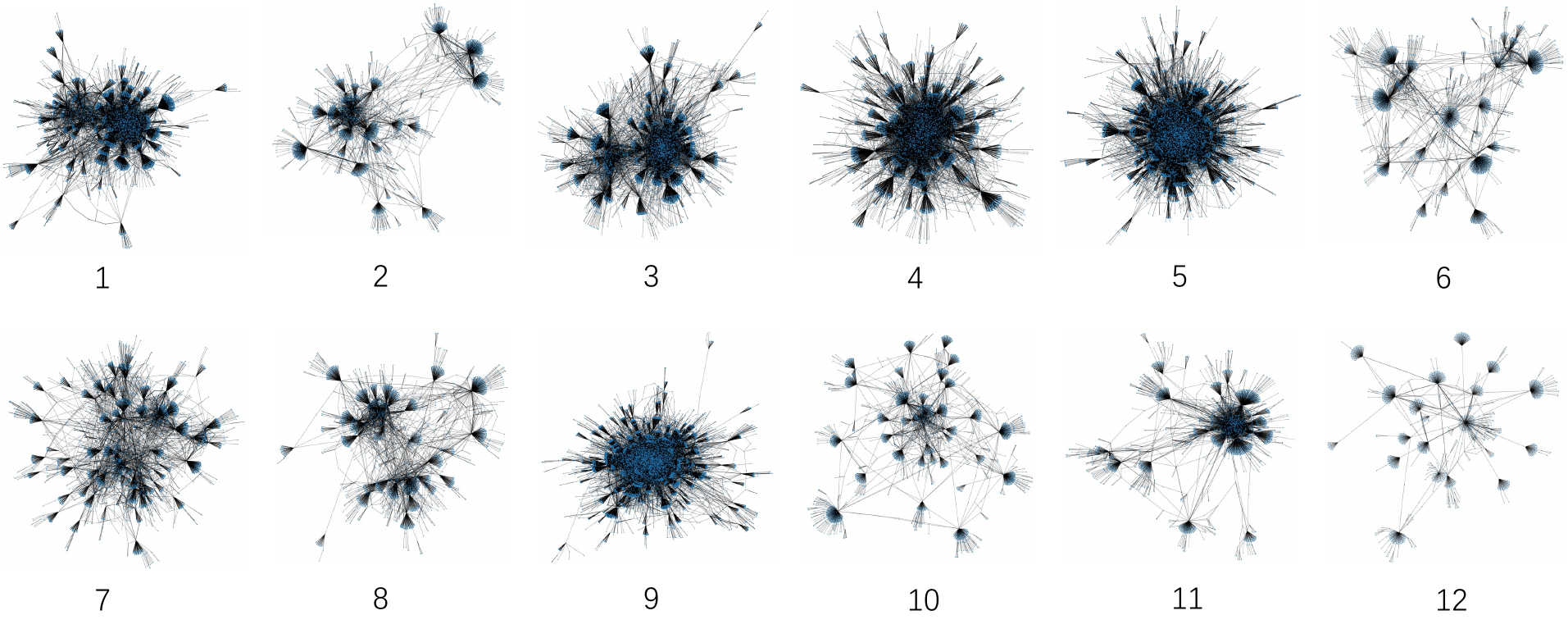}
    \caption{Visualization of user connectivity relationships in 12 communities.}
    \label{fig:user graph}
\end{figure}

We provide a detailed overview of the dataset used for EvoBot's learning and testing, including the number of users, tweets, and edges for each community, as shown in Table \ref{tab:comm stat}. And we visualize their structures as shown in \ref{fig:user graph}.
\begin{table}[ht]
    \centering
    \small
    
    \begin{tabular}{lccccc}
    \toprule
    \textbf{Comm} & \textbf{User} & \textbf{Bot} & \textbf{Edge} & \textbf{Tweet} & \textbf{ Language}\\
    \midrule
     1 & 4560 & 415 & 15137 & 266523 & ID \\
     2 & 1756 & 154 & 6346 & 100292 & EN \\
     3 & 3606 & 419 & 16214 & 336661 & IT, EN \\
     4 & 4269 & 747 & 15609 & 265188 & TR \\
     5 & 6923 & 628 & 23764 & 383878 & AR \\
     6 & 1254 & 253 & 4373 & 115758 & EN \\
     7 & 3399 & 633 & 10097 & 201882 & EN \\
     8 & 2004 & 273 & 5627 & 122147 & EN \\
     9 & 8347 & 992 & 26870 & 486288 & EN \\
     10 & 2187 & 190 & 5341 & 125544 & JA \\
     11 & 1085 & 256 & 6601 & 76615 & EN \\
     12 & 890 & 268 & 1898 & 45297 & EN \\
    \bottomrule
    \end{tabular}
    \caption{Summary of community data, including the number of users, bots, edges, tweets, and languages for each community.}
    \label{tab:comm stat}
\end{table}

To ensure EvoBot receives quality training data and avoids the influence of noisy or irrelevant information, we undertook a comprehensive process of data filtering and preprocessing. This process was divided into two key parts: the handling of Account Information and historical tweets, followed by the construction of a high-quality SFT dataset.

The first part focuses on the processing of Account Information and historical tweets. EvoBot aims to simulate individual users as accurately as possible, which requires embedding detailed user information into the prompt. Directly using raw data from the accounts would result in relatively low information density in the prompts. To address this, we employed GPT-4o to generate concise summaries of user information. The prompt is shown in Table \ref{tab:prompt}. Additionally, Figure \ref{fig:eg covid} provides an example of a summarized user profile.

The second part addresses the preparation of the SFT dataset. Since SFT requires high-quality data \cite{dong2023abilities}, we took steps to ensure the dataset met these standards. We removed incomplete sentences, excessive emoji use, and URL links from human tweets. Furthermore, we formatted the data output by structuring it in a sequential format, such as "1. \{Tweet 1\} \verb|\n| 2. \{Tweet 2\} \verb|\n|..." to maintain consistency and ensure EvoBot would learn effectively from clean and structured examples.

\section{Experimental Details} \label{appdix: training}
The pseudocode for EvoBot's learning is shown in Algorithm \ref{alg:evobot}, where \(\texttt{learning\_epochs}=4, N=1024, C=2\).

\begin{algorithm}[ht]
    \caption{EvoBot}
   \label{alg:evobot}
   \begin{algorithmic}
   \STATE \textbf{Initialize:}
   \STATE Detector $F^0=f^0$ by supervised learning on original dataset $D^0$
   \STATE EvoBot $\pi_{\theta}^0$ by SFT on Human data
   \FOR{$k$ in 1 to learning\_epochs}
   \STATE Initialize empty DPO dataset $D_{\text{DPO}}$
   \STATE Sample $N$ bot users with replacement
   \FOR{$i$ in 1 to $N$}
   \FOR{$c$ in 1 to $C$}
   \STATE Generate candidate response $T_{v_{b_i},c}$ by $\pi^{k-1}_{\theta}$
   \STATE Use $F^{k-1}$ to calculate the probability of $v_{b_i}$ being human with tweets $T_{v_{b_i},c}$
   \ENDFOR
   \STATE Get data tuple $(x^i, y^i_w, y_l^i)$, add it to $D_{\text{DPO}}$
   \ENDFOR
   \FOR{each bot $v_{b_i}$, where $i=1, 2, \dots, |\mathcal{B}|$}
   \STATE Generate new tweets $T'_{v_{b_i}}$
   \ENDFOR
   \STATE Replace all bot tweets in $D^{k-1}$ to get new dataset $D^k$
   \STATE Train classifier $f^k$ on $D^k$
   \STATE Update Detector: $F^{k}=\sum_{j=0}^{k}w^jf^j$
   \STATE Update EvoBot $\pi_{\theta}^k$ by DPO training on $D_{\text{DPO}}$
\ENDFOR
\end{algorithmic}
\end{algorithm}

\subsection{EvoBot} \label{appdx:generator}
The parameters used during EvoBot's training process, such as LoRA, SFT, DPO, and generation parameters (which are the same for the baseline LLM models), are provided in Tables \ref{tab:lora_config}, \ref{tab:sft training_config}, \ref{tab:dpo training_config}, and \ref{tab:generation training_config}, respectively. The prompt used in adversarial learning is shown in \ref{tab:prompt}. Figure \ref{fig:generator_example} gives an example.
\begin{table}[ht]
    \centering
    \begin{tabular}{l|c}
        \toprule
        \textbf{Parameter} & \textbf{Value} \\ 
        \midrule
        $r$ & 64 \\ 
        $\alpha$ (lora\_alpha) & 16 \\ 
        lora\_dropout & 0.1 \\ 
        task\_type & CAUSAL\_LM \\ 
        target\_modules & \{q,k,v,o\_proj\} \\ 
        \bottomrule
    \end{tabular}
    \caption{LoRA configuration parameters.}
    \label{tab:lora_config}
\end{table}

\begin{table}[ht]
    \centering
    \begin{tabular}{l|c}
        \toprule
        \textbf{Parameter} & \textbf{Value} \\ 
        \midrule
        per\_device\_train\_batch\_size & 2 \\ 
        per\_device\_eval\_batch\_size & 1 \\ 
        gradient\_accumulation\_steps & 32 \\ 
        bf16 & True \\ 
        learning\_rate & $2 \times 10^{-4}$ \\ 
        lr\_scheduler\_type & cosine \\ 
        warmup\_ratio & 0.1 \\ 
        max\_seq\_length & 2048 \\ 
        \bottomrule
    \end{tabular}
    \caption{SFT training configuration parameters.}
    \label{tab:sft training_config}
\end{table}

\begin{table}[ht]
    \centering
    \begin{tabular}{l|c}
        \toprule
        \textbf{Parameter} & \textbf{Value} \\ 
        \midrule
        $\beta$ & 0.2 \\
        per\_device\_train\_batch\_size & 1 \\ 
        per\_device\_eval\_batch\_size & 1 \\ 
        gradient\_accumulation\_steps & 32 \\ 
        bf16 & True \\ 
        max\_seq\_length & 2048 \\ 
        \bottomrule
    \end{tabular}
    \caption{DPO training configuration parameters.}
    \label{tab:dpo training_config}
\end{table}

\begin{table}[ht]
    \centering
    \begin{tabular}{l|c}
        \toprule
        \textbf{Parameter} & \textbf{Value} \\ 
        \midrule
        max\_length & 2048 \\
        do\_sample & True \\ 
        temperature & 0.7 \\ 
        repetition\_penalty & 1.3 \\ 
        top\_k & 50 \\ 
        top\_p & 0.6 \\ 
        \bottomrule
    \end{tabular}
    \caption{Generation parameters of all LLMs in our experiments.}
    \label{tab:generation training_config}
\end{table}

\begin{figure}
    \centering
    \includegraphics[width=0.98\columnwidth]{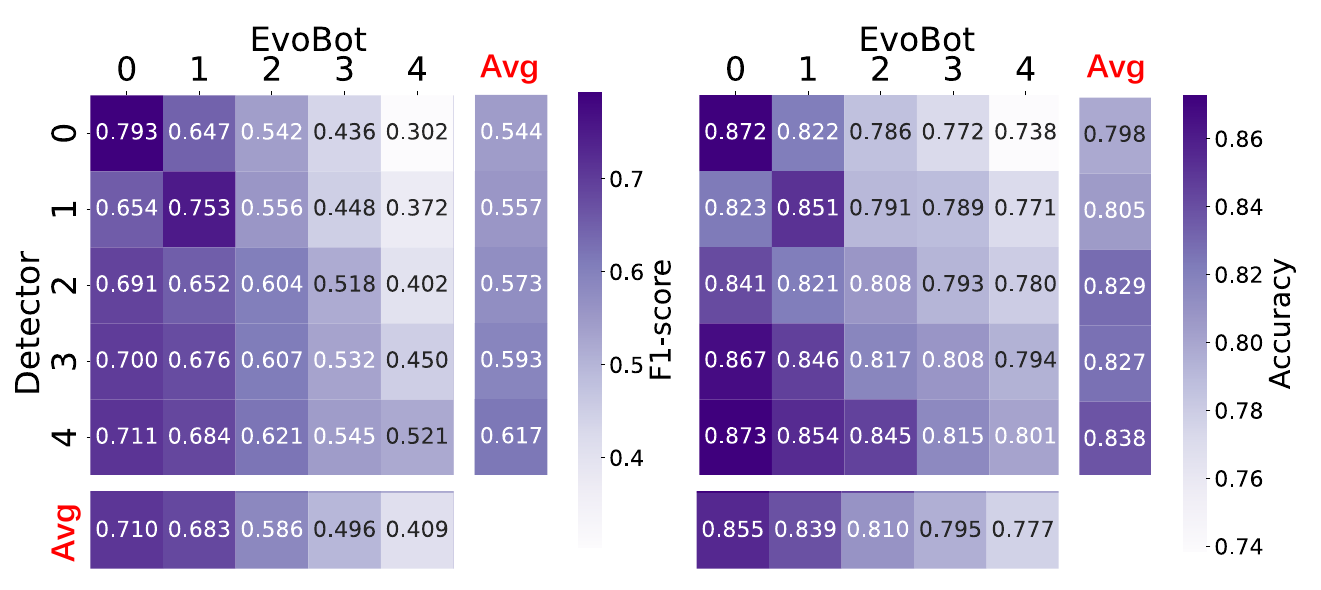}
    \caption{Classification performance (F1-score on the left, Accuracy on the right) across adversarial training iterations using Llama-3-8B-Instruct as the base model for EvoBot.}
    \label{fig:llama3 results}
\end{figure}

To explore the impact of the base Large Language Model on the co-evolutionary dynamics, we replicate the adversarial training experiment (as presented in Figure \ref{fig:adv results} of the main text) using Llama-3-8B-Instruct \cite{llama3modelcard} as the foundation for EvoBot. The classification performance, in terms of F1-score and Accuracy, across different iterations of EvoBot and the Detector is presented in Figure \ref{fig:llama3 results}. The results obtained with Llama-3 exhibit similar overall trends to those observed with the original Llama-2, suggesting that the adversarial learning framework remains effective in enhancing both the bot's evasiveness and the detector's robustness when a different underlying LLM is used.

Additionally, we have calculated the output diversity of each version of EvoBot. The results evaluate different EvoBot versions using language diversity metrics: Dist-1, Dist-2, Dist-3, and Shannon Entropy. Dist-1, Dist-2, and Dist-3 measure content diversity via n-gram overlap, with higher scores indicating greater variety. Shannon Entropy reflects text unpredictability, where higher values denote a more diverse and complex language model.

To provide a more comprehensive understanding of the internal quality of our generated content, we conduct an in-depth analysis focusing on the statistical significance of diversity, tweet length, and the evolution of stylistic features across training iterations.

We assess the statistical significance of EvoBot's diversity improvements across consecutive training versions using the Wilcoxon signed-rank test on the metrics reported in Table 2. The results, as shown in Table \ref{tab:diversity_significance}, confirm that the most significant gains in diversity occur during the initial transition from the base LLM (v0) to the first adversarial version (v1), with strong and highly significant p-values for all three Dist metrics. Subsequent iterations show more subtle, targeted refinements, with significant gains in specific metrics like Dist-2 and Dist-3 from v2 to v3, and a final small but significant gain in Dist-3 from v3 to v4. This progression from major, early gains to more refined, iterative improvements reflects a natural and effective learning trajectory for the generator.

\begin{table}[h]
\centering
\resizebox{0.5\textwidth}{!}{
\begin{tabular}{llcc}
\toprule
\textbf{Comparison} & \textbf{Metric} & \textbf{p-value} & \textbf{Effect Size (d)} \\
\midrule
\multirow{4}{*}{\textbf{v0$\rightarrow$v1}} & Dist-1 & $<$ 0.001 (***) & -0.968 \\
& Dist-2 & $<$ 0.001 (***) & -0.951 \\
& Dist-3 & $<$ 0.001 (***) & -0.509 \\
& SE & 0.019 (*) & 0.418 \\
\midrule
\multirow{4}{*}{\textbf{v1$\rightarrow$v2}} & Dist-1 & 0.003 (**) & -0.369 \\
& Dist-2 & 0.043 (*) & -0.350 \\
& Dist-3 & 0.768 (ns) & 0.099 \\
& SE & $<$ 0.001 (***) & 0.813 \\
\midrule
\multirow{4}{*}{\textbf{v2$\rightarrow$v3}} & Dist-1 & 0.685 (ns) & 0.072 \\
& Dist-2 & 0.024 (*) & 0.312 \\
& Dist-3 & $<$ 0.001 (***) & 0.685 \\
& SE & 0.951 (ns) & -0.032 \\
\midrule
\multirow{4}{*}{\textbf{v3$\rightarrow$v4}} & Dist-1 & 0.329 (ns) & -0.144 \\
& Dist-2 & 0.070 (ns) & -0.000 \\
& Dist-3 & 0.036 (*) & 0.287 \\
& SE & 0.672 (ns) & -0.010 \\
\bottomrule
\end{tabular}
}
\caption{Statistical significance of EvoBot's diversity improvements across versions. SE denotes Shannon Entropy. Significance levels: ns p $\ge$ 0.05; * p $<$ 0.05; ** p $<$ 0.01; *** p $<$ 0.001. Effect size (d): |d| $<$ 0.2 is negligible.}
\label{tab:diversity_significance}
\end{table}

To validate the human-likeness of EvoBot's generated content from a perceptual standpoint, we conduct a blind human evaluation. A total of 300 tweets are randomly sampled, comprising 100 each from real human users, our final EvoBot model, and a vanilla Llama 2 baseline. Five human annotators are tasked with classifying each tweet as either "Human" or "Bot." Concurrently, our final co-adapted Detector ($F^4$) evaluates the same set of tweets.

The results, presented in Table \ref{tab:human_validation}, show that human annotators find EvoBot's outputs significantly more difficult to distinguish from genuine human text compared to the Llama 2 baseline. Our EvoBot's tweets are misclassified as human at a much higher rate (62.0\%) than those from Llama 2 (48.1\%). This aligns with the Detector's performance, which also finds EvoBot more evasive, further corroborating its enhanced human-likeness.

To quantify the alignment between our automated detector and human perception, we calculate the Pearson correlation coefficient between the Detector's confidence scores and the human judgment scores (the proportion of annotators labeling a tweet as "Human"). This analysis yields a statistically significant positive correlation ($r = 0.78, p < 0.001$), demonstrating that our adversarial Detector is a reliable and scalable proxy for human evaluation of text authenticity.

\begin{table}[h]
\centering
\resizebox{0.5\textwidth}{!}{
\begin{tabular}{lcc}
\toprule
\textbf{Tweet Source} & \textbf{Human Acc} & \textbf{Detector Acc} \\
\midrule
\textbf{Llama 2} (as Bot) & 51.9\% & 48.2\% \\
\textbf{EvoBot} (as Bot) & 38.0\% & 35.3\% \\
\textbf{Human} (as Human) & 72.7\% & 87.0\% \\
\bottomrule
\end{tabular}
}
\caption{Comparison of human and Detector accuracy on a blind classification task. Lower accuracy for bot-generated tweets indicates higher perceived human-likeness.}
\label{tab:human_validation}
\end{table}

Furthermore, we conduct an analysis to measure the semantic similarity between the generated tweets and the user's real historical content. Using the sentence-transformers model `all-MiniLM-L6-v2`, we compute the cosine similarity for tweets generated during both the adversarial learning phase and the social simulation phase. As shown in Table \ref{tab:semantic_similarity}, EvoBot consistently achieves a higher similarity score to the target user's real tweets compared to the Llama baseline, indicating a better ability to capture individual user characteristics. The higher similarity during the learning phase reflects the SFT objective of replicating human tweets, while the lower score during simulation is expected, as the model's responses are guided by specific external events rather than general personal style.

\begin{table}[h]
\centering
\resizebox{0.5\textwidth}{!}{
\begin{tabular}{lcc}
\toprule
\textbf{Prompt Type} & \textbf{EvoBot Similarity} & \textbf{Llama 2 Similarity} \\
\midrule
Learning & \textbf{0.562} & 0.315 \\
Simulation & \textbf{0.285} & 0.251 \\
\bottomrule
\end{tabular}
}
\caption{Cosine similarity between generated tweets and a user's real historical tweets. Higher scores indicate stronger alignment with individual user characteristics.}
\label{tab:semantic_similarity}
\end{table}

An analysis of stylistic markers, such as the usage of emojis, hashtags, and mentions, across all EvoBot versions reveals a clear two-phase learning dynamic. As shown in Table \ref{tab:stylistic_markers}, the initial Supervised Fine-Tuning (SFT) phase induces a dramatic "correction" from the base LLM's behavior. The vanilla `Llama` model exhibits an overly frequent, bot-like usage of emojis and hashtags, which is significantly reduced in the SFT-trained EvoBot (v0). This demonstrates that SFT successfully grounds the model in a more natural, human-like baseline. Subsequently, during the iterative adversarial training phases (v1-v4), the usage rates of these markers do not converge to a static value but instead fluctuate within a lower, more human-like range. This suggests EvoBot learns a more nuanced and context-dependent application of these features rather than a simple, static rule.

\begin{table*}[h]
\centering
\begin{tabular}{lccccccc}
\toprule
\textbf{Metric} & \textbf{Llama} & \textbf{v0 (SFT)} & \textbf{v1} & \textbf{v2} & \textbf{v3} & \textbf{v4} & \textbf{Human} \\
\midrule
Emoji Usage & 52.2\% & 14.1\% & 18.4\% & 16.8\% & 12.6\% & 14.4\% & 17.1\% \\
Hashtag Usage & 69.5\% & 30.3\% & 12.4\% & 13.9\% & 4.9\% & 14.9\% & 18.6\% \\
Mention Usage & 15.0\% & 6.6\% & 4.8\% & 6.3\% & 5.4\% & 7.0\% & 10.8\% \\
\bottomrule
\end{tabular}
\caption{Stylistic feature usage rates for different EvoBot versions, compared to the vanilla LLM and real human users.}
\label{tab:stylistic_markers}
\end{table*}

\subsection{Detector}
The Detector model is a neural network designed for bot detection using Relational Graph Convolutional Networks (RGCN).  It takes four types of input features: user description, tweet content, numerical properties, and categorical properties, each passed through separate fully connected layers followed by LeakyReLU activation functions to generate embeddings. These embeddings are then concatenated and passed through another fully connected layer. The model utilizes two RGCNConv layers to perform graph convolution on the relational graph, followed by dropout for regularization. Finally, the output is passed through two more fully connected layers to produce the final prediction, which classifies the input into one of two categories (e.g., bot or human). The training parameters in adversarial learning are shown in Table \ref{tab:detector para}.

\begin{table}[ht]
\centering
\begin{tabular}{ll}
\toprule
\textbf{Parameter} & \textbf{Value} \\ \midrule
cat\_prop\_size & 3 \\ 
embedding\_dimension & 256 \\ 
dropout & 0.1 \\ 
lr & 1e-3 \\ 
weight\_decay & 0.1 \\ 
pretrain\_epochs & 120 \\ \bottomrule
\end{tabular}
\caption{Training parameters of the Detector.}
\label{tab:detector para}
\end{table}

Our previous ablation studies in Section 4.1.2 already show that both semantic content and relational graph structure are essential for the Detector's performance. To further probe the importance of the surface-level stylistic markers, we conduct an additional ablation study. We remove emojis, hashtags, and mentions from the generated text and evaluate our Detector's performance. The results, summarized in Table \ref{tab:detector_ablation_stylistic}, show that removing each of these features individually results in only a negligible and statistically insignificant decrease in the Detector's performance.

\begin{table}[h]
\centering
\resizebox{0.5\textwidth}{!}{
\begin{tabular}{lcc}
\toprule
\textbf{Condition} & \textbf{Acc} & \textbf{F1-score} \\
\midrule
\textbf{Original Text} & 0.892 $\pm$ 0.053 & 0.561 $\pm$ 0.042 \\
No Emojis & 0.890 $\pm$ 0.054 & 0.558 $\pm$ 0.043 \\
No Hashtags & 0.888 $\pm$ 0.054 & 0.555 $\pm$ 0.044 \\
No Mentions & 0.891 $\pm$ 0.053 & 0.560 $\pm$ 0.042 \\
\bottomrule
\end{tabular}
}
\caption{Detector performance after removing stylistic features from the input text. The negligible drop in performance suggests the detector relies on deeper semantic and relational cues.}
\label{tab:detector_ablation_stylistic}
\end{table}

This fine-grained analysis provides strong evidence that our Detector does not heavily rely on simple cues for its classification. It confirms that the detector leverages deeper semantic meaning from the text and relational signals from the social graph. This, in turn, demonstrates that EvoBot is compelled to achieve a more fundamental level of human-likeness to be successful, rather than just learning to manipulate superficial stylistic features.

\section{Social Simulation}

\subsection{Trigger News}
In the simulation of group opinion, two significant events are used: the COVID-19 pandemic and the Russian-Ukraine Conflict. These events are chosen due to their global impact and the intense discussions surrounding them on social media platforms.  Table \ref{tab:covid news} provides every trigger news of the COVID-19 event, while Table \ref{tab:ukraine news} outlines similar information for the Russian-Ukraine Conflict.

In the information spread simulation, only a subset of users are initially informed about the event: "The Los Angeles Rams clinched the 2022 Super Bowl championship with a thrilling 23-20 victory over the Cincinnati Bengals in Super Bowl LVI." 
\subsection{ABMs Model}
The \textbf{Bounded Confidence (BC)} model in opinion dynamics examines how individuals' opinions evolve through interactions constrained by a confidence threshold \(\epsilon\). Each individual \(i\) holds an opinion \(x_i(t) \in [0, 1]\), updated over time by interacting with another individual \(j\) only if \(|x_i(t) - x_j(t)| \leq \epsilon\). When this condition is met, opinions adjust symmetrically:
\[
x_i(t+1) = x_i(t) + \mu \cdot \left(x_j(t) - x_i(t)\right), 
\]

where \(\mu \in [0, 0.5]\) is the convergence parameter. Smaller \(\epsilon\) leads to opinion clusters, while larger \(\epsilon\) promotes consensus. Here, \(j\) is sampled from the users followed by \(i\), meaning that \(i\)'s opinion can be influenced by its following users.

The \textbf{Lorenz} model in opinion dynamics simulates how individual attitudes evolve through social interactions. Each agent \( i \) updates its attitude \( a_{it} \) at time \( t \) based on interactions with another agent \( j \). The update rule is:

\begin{equation*}
    \begin{split}
        \Delta a_{it} &= \alpha \cdot \text{pol}(a_{it}) \cdot \text{sim}(a_{it}, m_{jt}) \cdot \\
 & \left[ \theta \cdot (m_{jt} - a_{it}) + (1 - \theta) \cdot m_{jt} \right],
    \end{split}
\end{equation*}
where:
\begin{itemize}
    \item \( \alpha \): Susceptibility to change.
    \item \( \text{pol}(a_{it}) = \frac{M^2 - a_{it}^2}{M^2} \): Polarization factor.
    \item  \( \text{sim}(a_{it}, m_{jt}) = \frac{\lambda^k}{\lambda^k + |m_{jt} - a_{it}|^k} \): Similarity bias.
    \item  \( \theta \): Balance between assimilation (\( m_{jt} - a_{it} \)) and reinforcement (\( m_{jt} \)).
    \item \( m_{jt} = a_{jt} \): Message from agent \( j \).
\end{itemize}

Table \ref{tab: ABM model para} shows the parameters of them.

\begin{table}[ht]
\centering
\begin{tabular}{@{}lcc@{}}
\toprule
\textbf{Model} & \textbf{Parameter} & \textbf{Value} \\ \midrule
BC model       & $\mu$          & 0.8           \\
               & $\epsilon$ & 0.3           \\ \midrule
Lorenz model   & $\alpha$          & 0.1           \\
               & $\lambda$         & 2.0           \\
               & $k$               & 2.0           \\
               & $\theta$          & 0.5           \\ \bottomrule
\end{tabular}
\caption{Parameters for BC and Lorenz models.}
\label{tab: ABM model para}
\end{table}

\subsection{Group Opinion} \label{appendix: group opinion}
Figure \ref{fig:eg covid} illustrates an example of EvoBot's simulated tweet generation in response to a COVID-19 news topic. EvoBot’s tweet stands out by blending curiosity, relatability, and a casual tone. Unlike GPT and Llama, which offer more formal and neutral responses, EvoBot incorporates personal touches like "I just read" and humor ("haha jk"), making it feel more human-like and engaging. It also showcases empathy with phrases like "my fellow humans," reflecting a thoughtful and personal approach to the topic. While GPT maintains a professional tone and Llama adds a more action-oriented perspective, EvoBot excels in creating a conversational, approachable atmosphere that resonates with users.

\begin{figure}[ht]
\centering
\subfloat   %
  {
\label{fig:covid}\includegraphics[width=0.98\columnwidth]{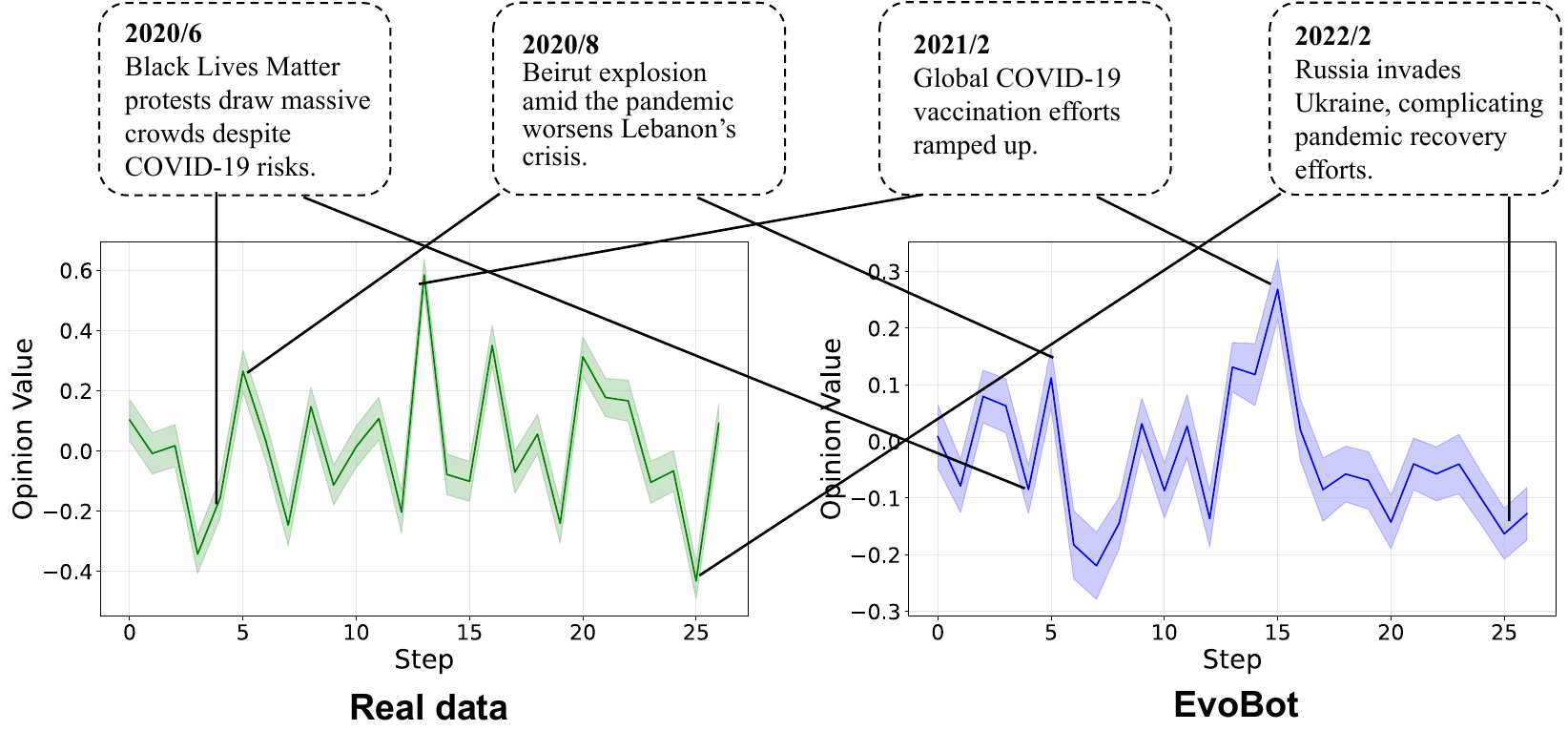} 
  }

  \caption{Comparison of real-world opinion dynamics and EvoBot-generated opinion dynamics regarding COVID-19.}
  \label{fig:covid curve}
\end{figure}

Figure \ref{fig:covid curve} compares real-world opinion dynamics with EvoBot-generated opinion dynamics regarding COVID-19. The left panel shows actual public opinion over time, highlighting significant events such as the Black Lives Matter protests in June 2020, the Beirut explosion in August 2020, and the global COVID-19 vaccination efforts in February 2021. The right panel presents EvoBot’s simulated opinion dynamics, reflecting similar fluctuations in response to these events. 

\begin{figure}[ht]
\centering
\subfloat   %
  {
      \label{fig:ukraine}\includegraphics[width=0.98\columnwidth]{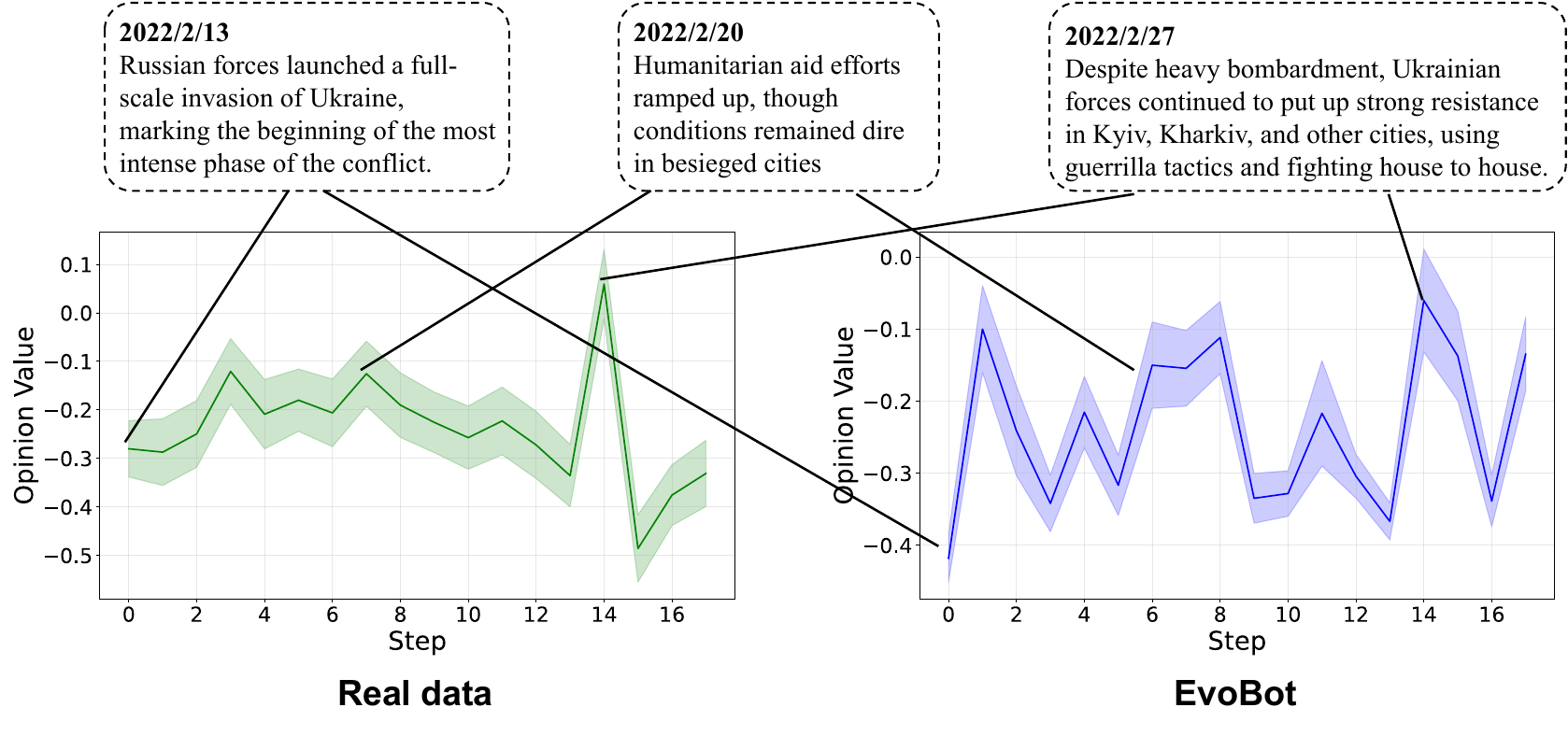} 
  }

  \caption{Comparison of real-world opinion dynamics and EvoBot-generated opinion dynamics during the Russia-Ukraine Conflict.}
  \label{fig:ukraine curve}
\end{figure}

\begin{figure*}[ht]
    \centering
    \includegraphics[width=0.98\linewidth]{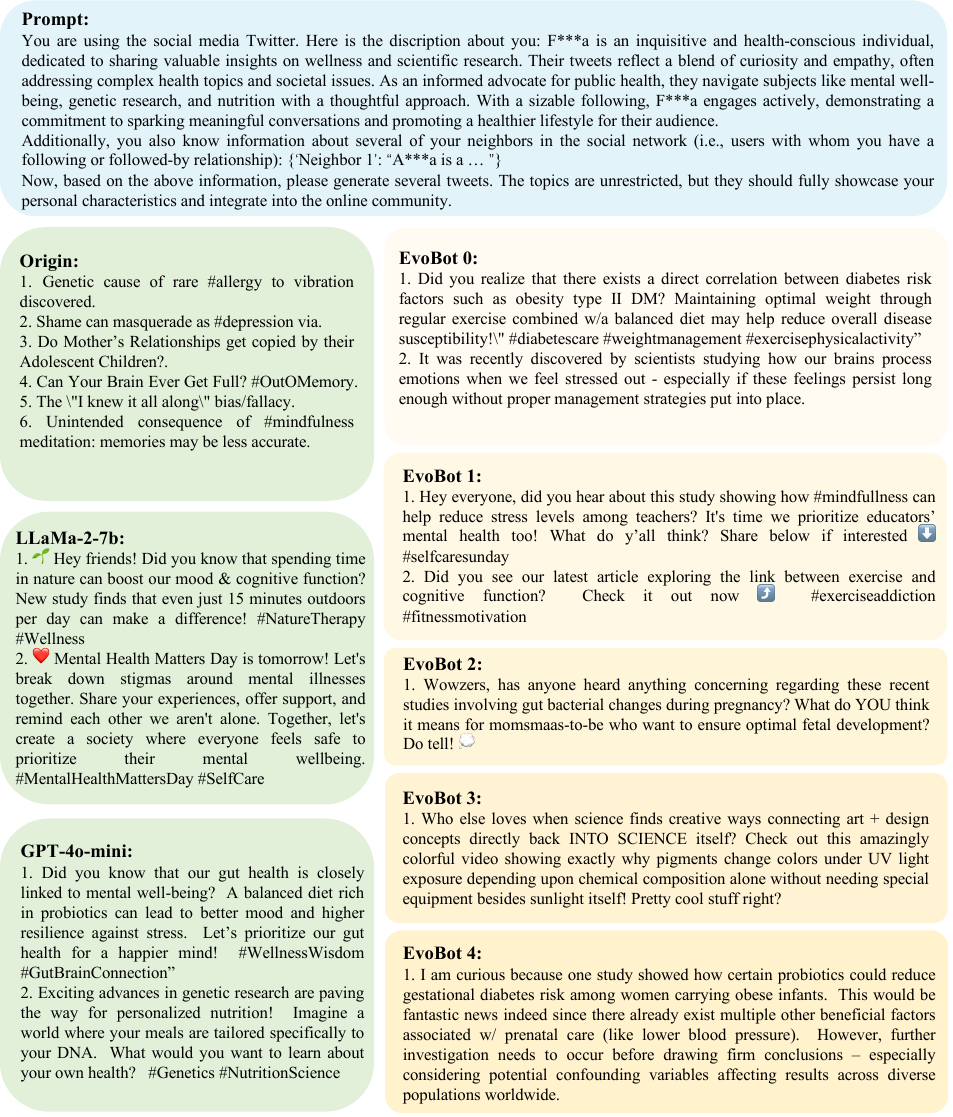}
    \caption{The tweet generation of different versions of EvoBot, Llama2-7b, and GPT-4o-mini based on a prompt for a health-conscious individual.}
    \label{fig:generator_example}
\end{figure*}

\begin{figure*}[ht]
    \centering
    \includegraphics[width=0.98\textwidth]{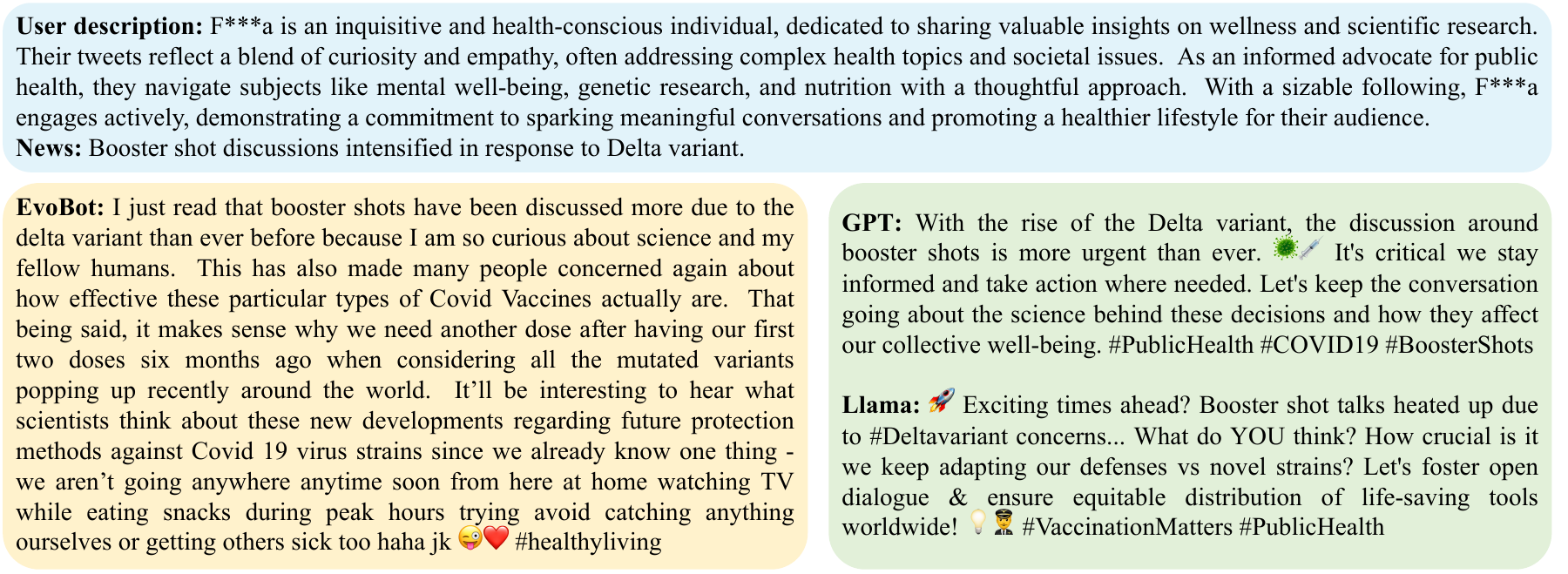}
    \caption{Example of the simulated tweet generation in response to a COVID-19-related news topic.}
    \label{fig:eg covid}
\end{figure*}

Figure \ref{fig:ukraine curve} compares real-world opinion dynamics with EvoBot-generated opinion dynamics during the Russia-Ukraine Conflict. The left panel displays real public opinion data over time, highlighting key events such as the full-scale Russian invasion of Ukraine on February 13, 2022, the ramping up of humanitarian aid efforts on February 20, 2022, and continued Ukrainian resistance despite heavy bombardment on February 27, 2022. The right panel shows EvoBot’s simulated opinion dynamics, which reflect similar trends and fluctuations in response to these events. 

EvoBot’s simulation demonstrates its capability to replicate real-world opinion shifts in a context-sensitive manner, showcasing its effectiveness in mimicking public sentiment during key global events.

Figure \ref{fig:abm results} presents the results of the BC and Lorenz models in group opinion simulations for two major global events: COVID-19 and the Russia-Ukraine Conflict. Figure \ref{fig:bc covid} shows the BC model's dynamics in the context of COVID-19, where the opinion values rapidly stabilize into distinct clusters after a few steps, reflecting the polarization of opinions within the group. Figure \ref{fig:bc ru} displays the BC model applied to the Russia-Ukraine Conflict, where the opinions also converge but with a faster decline in opinion diversity.

Figure \ref{fig:lorenz covid} and \ref{fig:lorenz ru} illustrate the behavior of the Lorenz model in the same two contexts. In \ref{fig:lorenz covid}, the Lorenz model applied to COVID-19 shows more continuous oscillations in the opinion values, with groups fluctuating around their final states. In \ref{fig:lorenz ru}, the Lorenz model in the Russia-Ukraine Conflict presents more rapid opinion convergence.

\begin{table*}[ht]
    \centering
    \renewcommand\tabcolsep{5.2pt}
\renewcommand\arraystretch{1.2}
    \begin{tabular}{c|p{12cm}}
    \toprule
       Summarization  & Summarize user:
Generate a character description based on the following user information: \newline
Name: \{...\} \newline
Location: \{...\} \newline
Description: \{...\}\newline
Account Created: \{...\} \newline
Followers Count: \{...\} \newline
Following Count: \{...\} \newline
Tweet Count: \{...\} \newline
Sample of Previous Posts: \{...\} \newline
Please include inferred personality traits and a summary of their Twitter activity. Only return a short description and other words are NOT allowed. Avoid repeating the observation in the summary.
 \\
 \hline
       EvoBot in Learning  &  Now you are acting as an agent named \{agent\_name\} in the social media Twitter. Here are some information:\newline
(1) The agent's description: \{role\_description\}; \newline
(2) Current time is \{current\_time\};\newline
(3) The news you got is \{trigger\_news\};\newline
(4) The events that occurred in the past are \{past\_event\};\newline
(5) The twitter page you can see is \{tweet\_page\}.\newline
Use the information to assess if the user is interested in this news. If the user is, compose a tweet expressing an opinion. If not, write a random tweet. \\
\hline
EvoBot in Simulation & You are using the social media Twitter. Here is the description about you: \{USER\_SUMMARY\}.
Additionally, you also know information about several of your neighbors in the social network (i.e., users with whom you have a following or followed-by relationship): \{NEIGHBORS\_SUMMARY\}
Now, based on the above information, please generate several tweets. The topics are unrestricted, but they should fully showcase your personal characteristics and integrate into the online community. \\
       \bottomrule
    \end{tabular}
    \caption{Prompt templates used in our experiments.}
    \label{tab:prompt}
\end{table*}

\begin{figure*}[ht]
    \centering
    \includegraphics[width=0.98\textwidth]{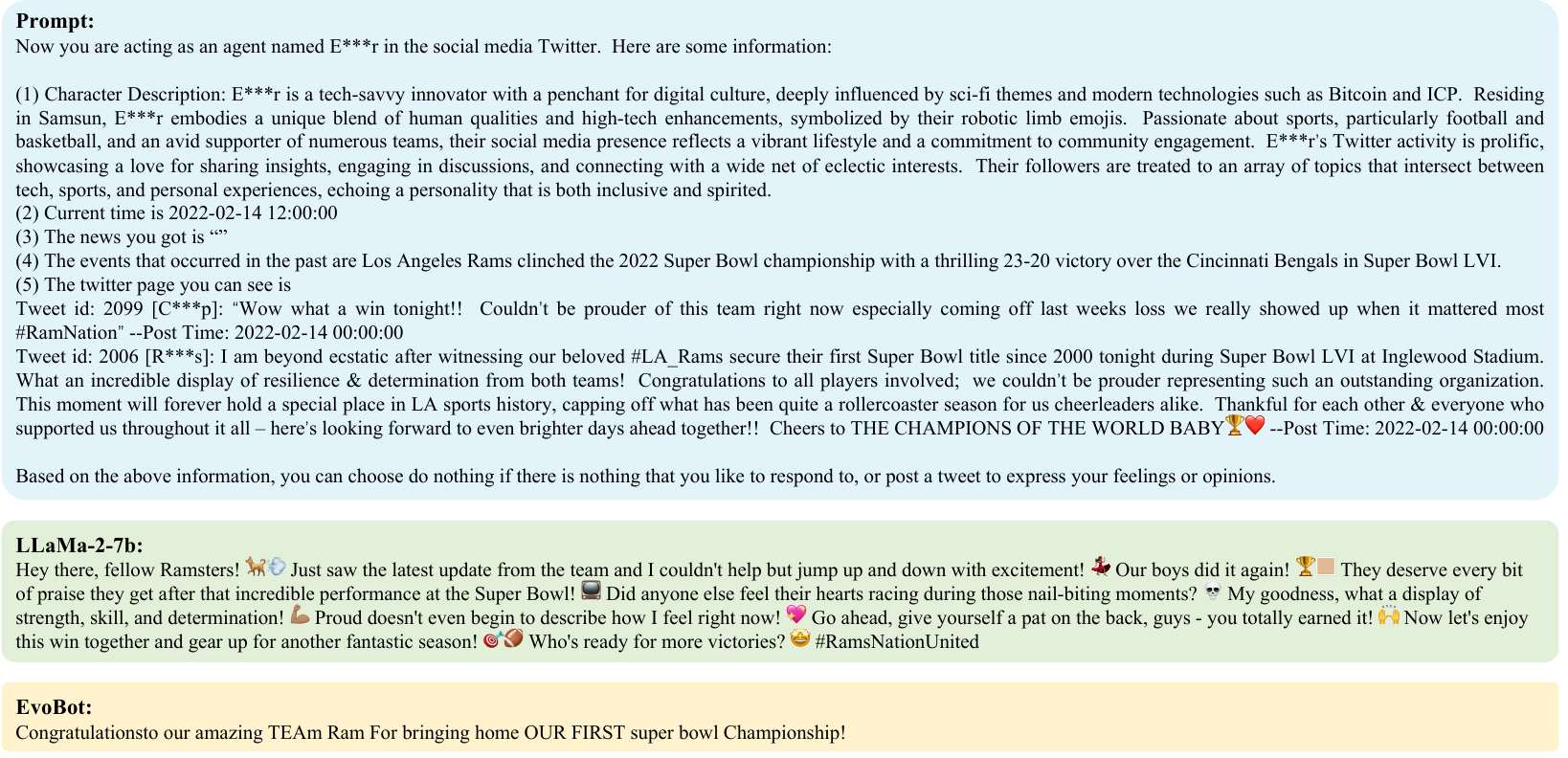}
    \caption{Example of tweet generation during a round of information spread simulation.}
    \label{fig:eg info spread}
\end{figure*}

\begin{figure*}[ht]
\centering
\subfloat[BC in COVID-19]
  {
      \label{fig:bc covid}\includegraphics[width=0.98\columnwidth]{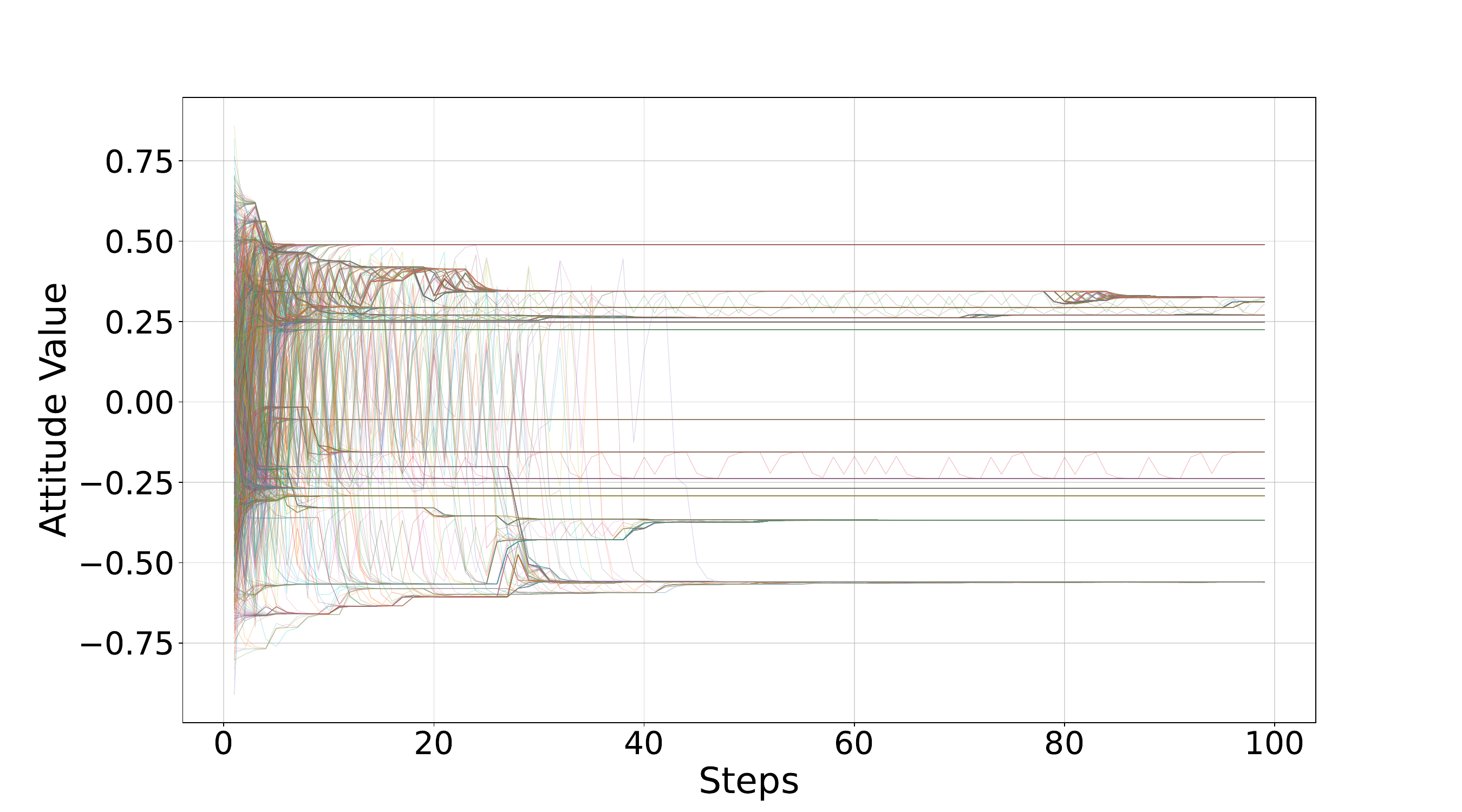} 
  } 
  \subfloat[BC in Russia-Ukraine Conflict]
  {\label{fig:bc ru}\includegraphics[width=0.98\columnwidth]{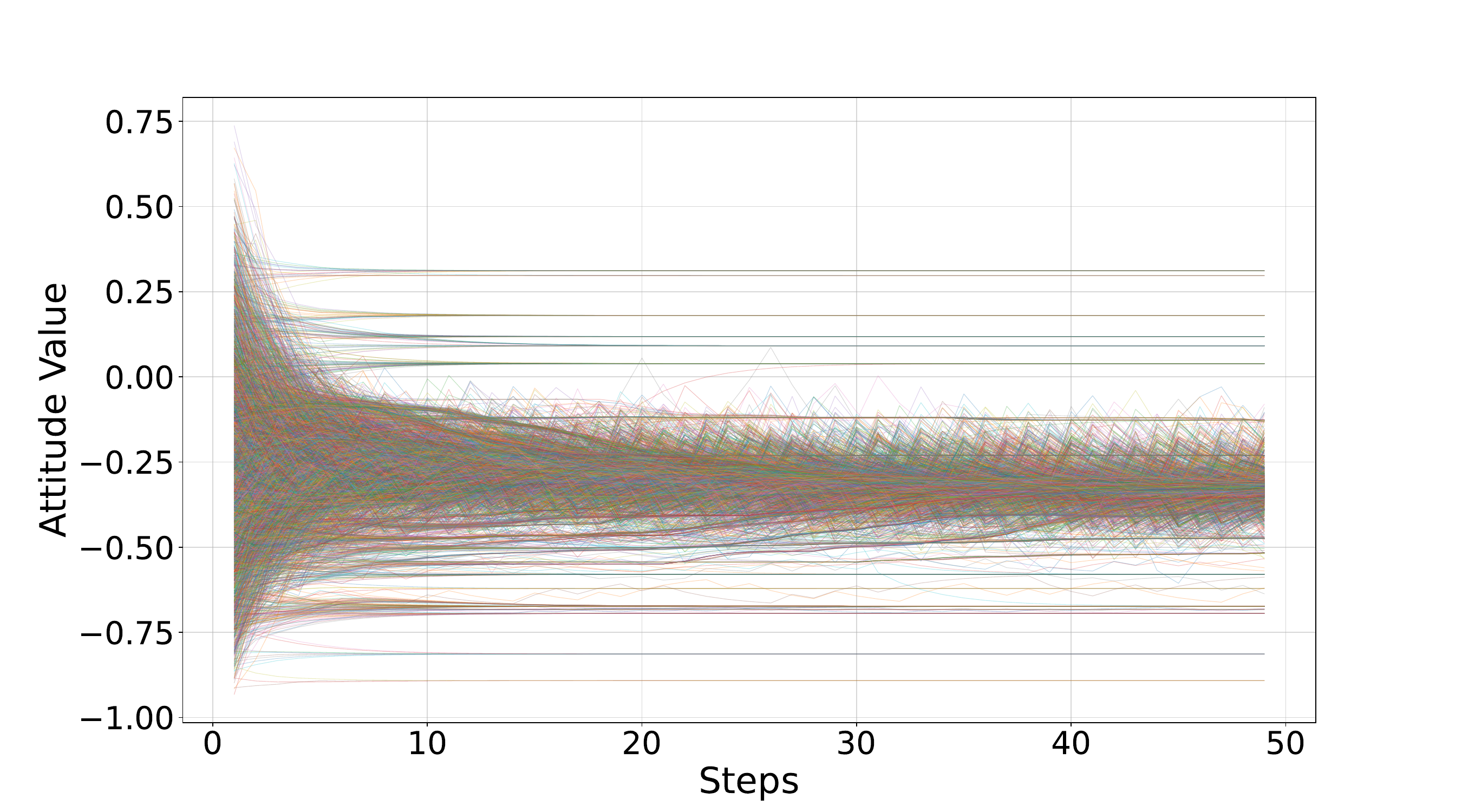} } \\
  \subfloat[Lorenz in COVID-19]
  { \label{fig:lorenz covid}\includegraphics[width=0.98\columnwidth]{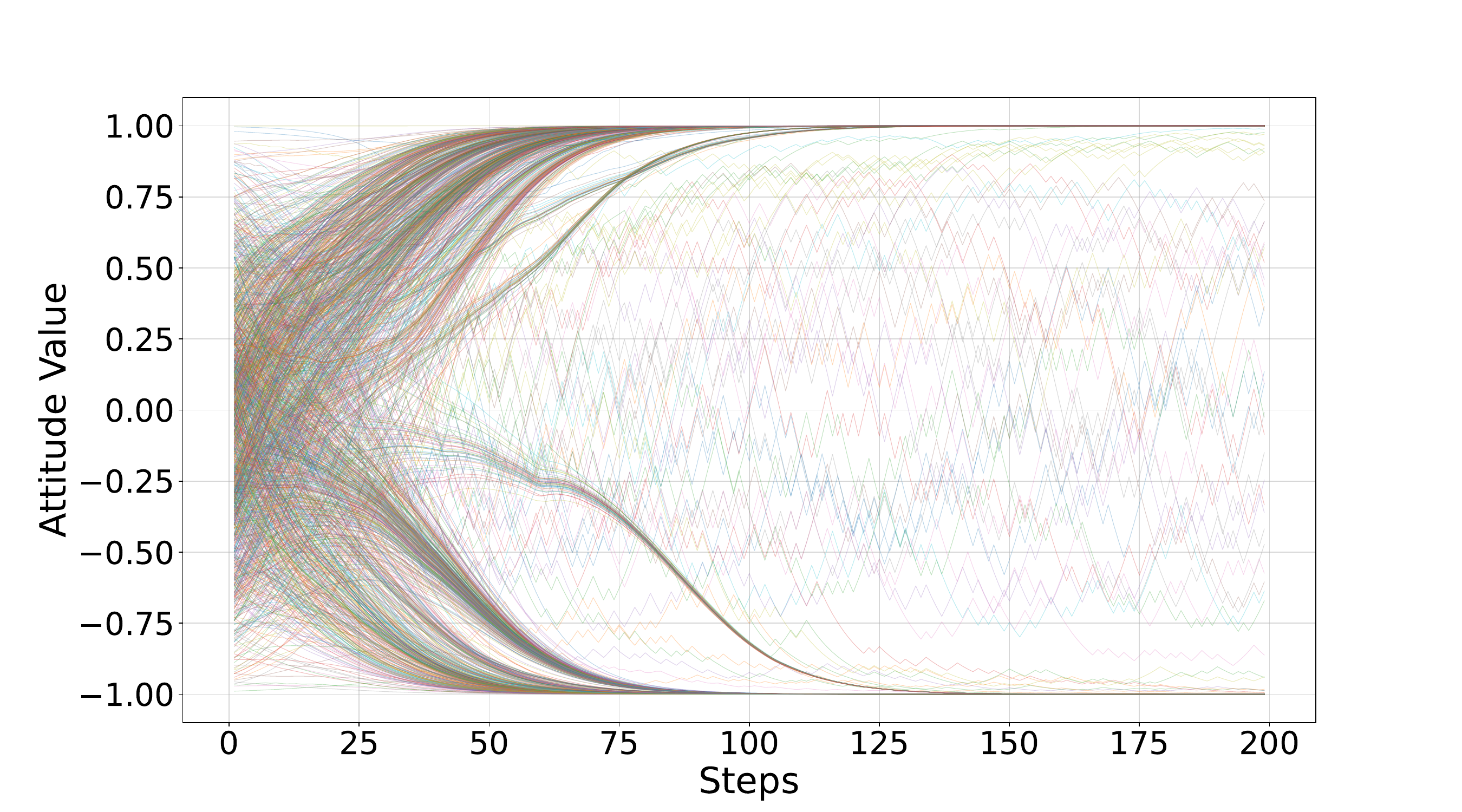} }
\subfloat[Lorenz in Russia-Ukraine Conflict]
  { \label{fig:lorenz ru}\includegraphics[width=0.98\columnwidth]{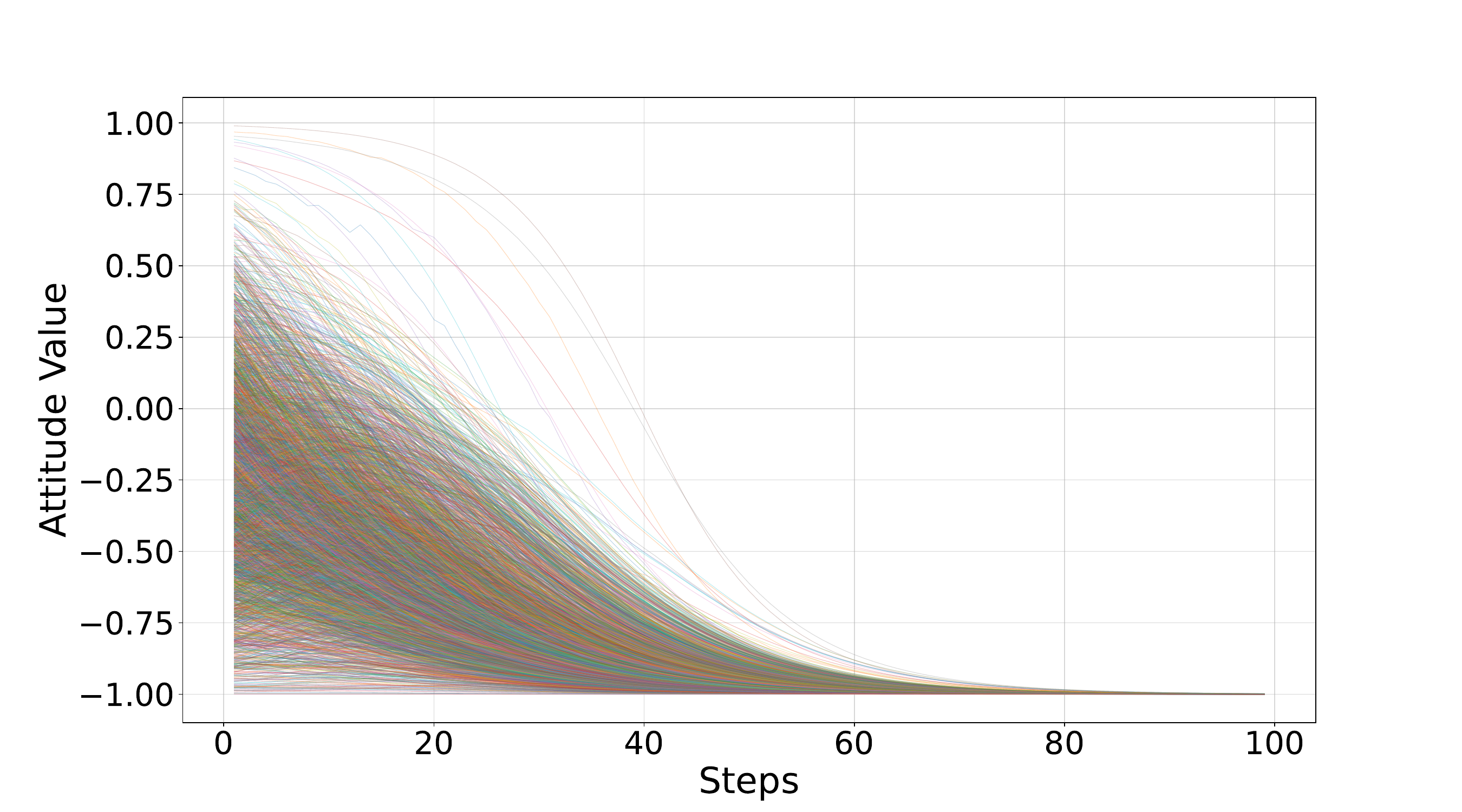} }
  \caption{Results of BC and Lorenz models in group opinion simulation}
  \label{fig:abm results}
\end{figure*}

\begin{table*}[ht]
    \centering
    \renewcommand\tabcolsep{5.2pt}
\renewcommand\arraystretch{1.2}
\resizebox{0.8\textwidth}{!}{
    \begin{tabular}{c|p{12cm}}
    \toprule
      Time  & News \\
     \hline
 2020/01 & WHO issues a global warning about a new coronavirus spreading in Wuhan, China.\\
 2020/02 & Diamond Princess cruise ship outbreak highlights virus transmissibility.\\
 2020/03 & WHO declares COVID-19 a pandemic.\\
 2020/04 & Mass graves in New York City for unclaimed COVID-19 victims.\\
 2020/05 & Anti-lockdown protests erupt in the U.S. and Europe.\\
 2020/06 & Black Lives Matter protests draw massive crowds despite COVID-19 risks.\\
 2020/07 & Surge in cases following Independence Day gatherings in the U.S.\\
 2020/08 & Beirut explosion amid the pandemic worsens Lebanon's crisis.\\
 2020/09 & India reports over 90,000 daily cases, marking a global peak.\\
 2020/10 & White House COVID-19 outbreak infects President Trump.\\
 2020/11 & Europe's second wave leads to renewed lockdowns.\\
 2020/12 & New COVID-19 variant discovered in the UK.\\
 2021/01 & U.S. Capitol riot amid record COVID-19 deaths.\\
 2021/02 & Global COVID-19 vaccination efforts ramped up.\\
 2021/03 & Brazil's healthcare system collapses amid rising cases.\\
 2021/04 & India experiences oxygen shortages during the second wave.\\
 2021/05 & Tokyo Olympics proceed without spectators.\\
 2021/06 & Delta variant spreads rapidly worldwide.\\
 2021/07 & The highly transmissible Delta variant caused a rapid increase in COVID-19 cases worldwide. Hospitals in many countries, including the U.S., India, and Indonesia, were overwhelmed, leading to rising fears about the variant’s impact on vaccine effectiveness.\\
 2021/08 & Reports highlighted the stark inequality in vaccine distribution, with wealthy countries administering booster shots while poorer nations struggled to vaccinate even frontline workers. This fueled global criticism and fear of prolonged pandemic impacts.\\
 2021/09 & The World Health Organization (WHO) classified the Mu variant (B.1.621) as a “variant of interest.” Concerns grew about its potential to evade immunity from prior infections or vaccinations, adding to global anxiety.\\
 2021/10 & WHO warns of slow vaccination rates in Africa.\\
 2021/11 & Omicron variant identified in South Africa.\\
 2021/12 & Omicron-driven surge overwhelms global healthcare systems.\\
 2022/01 & COVID-19 cases reach record highs globally.\\
 2022/02 & Russia invades Ukraine, complicating pandemic recovery efforts.\\
 2022/03 & Shanghai enters strict lockdown amid China's zero-COVID policy. \\
\bottomrule 
    \end{tabular}}
    \caption{Key events related to the COVID-19 pandemic, covering major global developments from the early stages of the outbreak through the challenges of new variants and the ongoing efforts for pandemic recovery.}
    \label{tab:covid news}
\end{table*}

\begin{table*}[ht]
    \centering
    \renewcommand\tabcolsep{5.2pt}
\renewcommand\arraystretch{1.2}
\resizebox{0.8\textwidth}{!}{
    \begin{tabular}{c|p{14cm}}
    \toprule
Time  & News \\
     \hline
2/13 &  Russian forces launched a full-scale invasion of Ukraine, marking the beginning of the most intense phase of the conflict. The attack included airstrikes, ground invasions, and naval assaults targeting major Ukrainian cities, including Kyiv, Kharkiv, and Odessa.\\
2/14 &  Ukrainian President Volodymyr Zelenskyy rejected an offer of evacuation from the U.S., stating that he needed ammunition, not a ride. Ukrainian forces fiercely resisted Russian advances despite being outnumbered.\\
2/15 &  Western countries, including the U.S., European Union, and the UK, imposed heavy sanctions on Russia, targeting banks, businesses, and prominent individuals. NATO countries began sending weapons and supplies to Ukraine.\\
2/16 &  Russian forces took control of the Chernobyl nuclear power plant, which had been the site of a catastrophic nuclear disaster in 1986. This raised fears of a nuclear incident amid the ongoing conflict.\\
2/17 &  The UN held an emergency session in response to Russia's invasion, with many countries condemning the aggression. Russia vetoed a resolution that would have demanded a ceasefire and withdrawal of forces from Ukraine.\\
2/18 &  Russian troops moved closer to Kyiv, Ukraine's capital, while intensifying their assault on cities in eastern Ukraine. Meanwhile, Russia announced it was placing its nuclear forces on alert.\\
2/19 &  Ukraine formally applied for European Union membership, emphasizing its desire to align more closely with Western Europe and away from Russian influence.\\
2/20 &  Thousands of Ukrainians fled westward to neighboring countries, especially Poland, as the war caused a massive refugee crisis. Humanitarian aid efforts ramped up, though conditions remained dire in besieged cities.\\
2/21 &  Russian forces continued to move toward Kyiv, and the city became a focal point of fierce fighting. Ukrainian President Zelenskyy remained in Kyiv, despite calls for his evacuation.\\
2/22 &  Ukrainian cities, including Mariupol, faced severe bombardment. Reports began emerging of significant civilian casualties and destruction due to Russian artillery and airstrikes.\\
2/23 &  Russian troops effectively encircled Mariupol, a port city in southern Ukraine, cutting off supplies and trapping thousands of civilians.\\
2/24 &  NATO leaders met to discuss increased defense aid for Ukraine, while the EU announced new sanctions against Russia, including restrictions on its access to financial systems and technology.\\
2/25 &  The international community, including the UN, continued to condemn Russia's actions. Reports of Russian war crimes, including targeting civilians and hospitals, emerged from various parts of Ukraine.\\
2/26 &  Humanitarian aid convoys attempted to reach the city, but Russian forces blocked routes, continuing their siege. Meanwhile, the UN confirmed over 2 million refugees had fled Ukraine.\\
2/27 &  Despite heavy bombardment, Ukrainian forces continued to put up strong resistance in Kyiv, Kharkiv, and other cities, using guerrilla tactics and fighting house to house.\\
2/28 &  The UN General Assembly passed a resolution demanding Russia cease its invasion of Ukraine, with a significant majority of countries voting in favor, though Russia and a few allies opposed it.\\
3/1  &   Russian troops captured large parts of southern Ukraine, including the city of Kherson, which became the first major city to fall under Russian control.\\
3/2  &   Russia continued its military advance, focusing on strategic locations like Mariupol, which remained besieged, while fighting continued on multiple fronts, especially in the Donbas region. \\
\bottomrule 
    \end{tabular}
    }
    \caption{Timeline of key events during the early stages of Russian-Ukraine Conflict in 2022.}
    \label{tab:ukraine news}
\end{table*}

To rigorously evaluate the performance of EvoBot in group opinion simulations, we conducted a Mann-Whitney U test comparing EvoBot's average bias ($\Delta_{Bias}$) and average diversity difference ($\Delta_{Div}$) against all baseline models. The results in Table \ref{tab:simulation_significance} confirm that EvoBot's superior performance is statistically significant in the majority of cases. Notably, EvoBot demonstrates highly significant improvements ($p < 0.001$) over traditional ABMs (BC and Lorenz), especially in replicating opinion diversity, and consistently outperforms LLM baselines like Llama and GPT.

\begin{table*}[ht]
\centering
\begin{tabular}{lcccc}
\toprule
\textbf{Comparison} & \textbf{COVID-19 $\Delta_{Bias}$} & \textbf{COVID-19 $\Delta_{Div}$} & \textbf{RU-UA $\Delta_{Bias}$} & \textbf{RU-UA $\Delta_{Div}$} \\
\midrule
\textbf{EvoBot vs. BC} & $<$ 0.05 (*) & $<$ 0.001 (***) & 0.412 (ns) & $<$ 0.001 (***) \\
\textbf{EvoBot vs. Lorenz} & $<$ 0.01 (**) & $<$ 0.001 (***) & $<$ 0.001 (***) & $<$ 0.001 (***) \\
\textbf{EvoBot vs. Llama2} & $<$ 0.01 (**) & $<$ 0.001 (***) & $<$ 0.001 (***) & $<$ 0.01 (**) \\
\textbf{EvoBot vs. GPT-4o} & 0.185 (ns) & $<$ 0.05 (*) & $<$ 0.05 (*) & $<$ 0.05 (*) \\
\bottomrule
\end{tabular}
\caption{Statistical significance of EvoBot's simulation performance against baselines using the Mann-Whitney U test. Significance levels are: ns $p \ge 0.05$; * $p < 0.05$; ** $p < 0.01$; *** $p < 0.001$.}
\label{tab:simulation_significance}
\end{table*}

We further address the stability and robustness of our simulation results. Our experimental design already accounts for varying network structures by running simulations on three distinct communities (comm2, comm5, and comm7), which exhibit notable topological differences (as detailed in Figure 5 and Table 6). The consistent outperformance of EvoBot across these communities demonstrates the robustness of its learned behaviors.

To quantify the stability of our results against stochasticity, we repeated each key simulation experiment three times and calculated the mean and standard deviation for our primary metrics, $\Delta_{Bias}$ and $\Delta_{Div}$. As shown in Table \ref{tab:simulation_stability}, the standard deviations are consistently small across all models. This indicates that the conclusions drawn, particularly EvoBot's superior performance, are reliable and not a result of random variation.

\begin{table*}[ht]
\centering
\begin{tabular}{lcccc}
\toprule
\textbf{Simulation} & \textbf{Metric} & \textbf{Llama2-7b} & \textbf{GPT-4o-mini} & \textbf{EvoBot (Ours)} \\
\midrule
\multirow{2}{*}{\textbf{COVID-19}} & $\Delta_{Bias}$ & 0.096 $\pm$ 0.005 & 0.083 $\pm$ 0.004 & \textbf{0.074 $\pm$ 0.003} \\
& $\Delta_{Div}$ & 0.107 $\pm$ 0.006 & 0.085 $\pm$ 0.005 & \textbf{0.054 $\pm$ 0.004} \\
\midrule
\multirow{2}{*}{\textbf{RU-UA Conflict}} & $\Delta_{Bias}$ & 0.205 $\pm$ 0.010 & 0.132 $\pm$ 0.008 & \textbf{0.103 $\pm$ 0.006} \\
& $\Delta_{Div}$ & 0.261 $\pm$ 0.012 & 0.240 $\pm$ 0.011 & \textbf{0.197 $\pm$ 0.009} \\
\bottomrule
\end{tabular}
\caption{Stability of key simulation metrics across three runs, showing mean and standard deviation.}
\label{tab:simulation_stability}
\end{table*}

\subsection{Information Spread}
Figure \ref{fig:eg info spread} shows an example of tweet generation during a round of information spread simulation, highlighting the role of EvoBot in producing concise and direct responses. While both Llama and EvoBot generate content reflecting excitement and community engagement, EvoBot’s response stands out for its brevity and focused messaging. This advantage makes EvoBot particularly effective for information spread, as shorter, more direct messages are often more easily disseminated and shared within a social network, enhancing the speed and reach of the information flow.

\section{AI Assistants Usage}
The generative AI tools, specifically ChatGPT and Copilot, are used during the research and writing process. ChatGPT assists with language refinement (e.g., paraphrasing and grammar correction), while Copilot is used for code-related tasks. Neither tool generates novel ideas, and all outputs are reviewed and edited by the authors.
\end{document}